\documentclass[11pt]{article}
\usepackage[margin=1in]{geometry}
\usepackage[utf8]{inputenc}
\usepackage[numbers,sort&compress]{natbib}
\usepackage[T1]{fontenc}
\usepackage{mathptmx}
\usepackage{amsmath,amssymb}
\usepackage{graphicx}
\usepackage{booktabs}
\usepackage{array}
\usepackage{caption}
\usepackage{microtype}
\usepackage{tikz}
\usepackage{xcolor}
\usepackage[colorlinks=true,linkcolor=blue!50!black,citecolor=blue!50!black,
            urlcolor=blue!50!black]{hyperref}

\usetikzlibrary{positioning}

\newcommand{\Ident}{\mathbb{I}}
\newcommand{\Ch}{\mathcal{C}}
\newcommand{\adel}{a_{\rm del}}
\newcommand{\Tr}{^{\mathrm{T}}} 
\newif\ifhighlightfixes
\highlightfixestrue
\definecolor{fixcol}{RGB}{0,90,180}
\ifhighlightfixes

\else

\fi

\title{Physics-Informed Drift Diagnosis for\\ Laser-Plasma Accelerator Operations}
\author{
Ou Labun\thanks{Corresponding author: \texttt{ozlabun@utexas.edu}},\;
Calin Hojbota,\;
Mara Klebonas,\;
Mike Downer,\;
Rafal Zgadzaj\\
\emph{Center for High Energy Density Science, The University of Texas at Austin, Austin, TX 78712}\\[4pt]
Phil Franke,\;
Reinier van Mourik,\;
Lance Labun\\
\emph{Tau Systems Inc, Austin, Texas, 78701, USA}
}
\date{\today}

\begin{document}

\maketitle

\begin{abstract}
Laser-plasma accelerators (LPAs) sustain accelerating gradients of order $100\,\mathrm{GV/m}$, but routine operation remains difficult: electron beam metrics drift over an operating shift, and the root physical cause is often invisible to the available diagnostics. We formulate LPA operation as a latent state-space model in which three effective interaction-point variables, the normalized laser amplitude $a_0$, the normalized plasma electron density $\tilde n_e$ and the residual pulse chirp $\mathcal{C}$, are inferred from routine electron beam observations by an extended Kalman filter. The emission model, which maps the latent state to the diagnostics, is kept structurally separate from the {transition} model, which describes how the latent state evolves between shots. The separation supports diagnosis in two stages, one asking which latent variable moved and one asking what moved it. The implemented emission model is a toy model, yielding an expected performance in line with current facilities and using 3D blow-out regime dependencies where relevant. We conduct synthetic sessions to test the effectiveness of the detection and attribution protocols, finding that attribution is limited by excitation rather than by shot count or diagnostic resolution.  Because the construction needs only a set of physical latent variables, an emission model and a family of hardware-derived transition models, it transfers to other drift-prone subsystems. We argue that the accuracy of the whole procedure is limited by the emission model rather than by the inference method.
\end{abstract}

\section{Introduction}
\label{sec:intro}

A laser-plasma accelerator (LPA) uses the plasma wave driven by an intense, ultrashort laser pulse to accelerate electrons. Because the accelerating structure is an ionized plasma rather than a metallic cavity, it is not limited by material breakdown, and gradients three orders of magnitude above those of radio-frequency structures are routine~\cite{esarey2009,downer2018diagnostics}. LPAs have delivered multi-GeV electron bunches~\cite{gonsalves2019, aniculaesei2024acceleration} and have driven a free-electron laser~\cite{wang2021,barber2025greater}. An electron beam that would require tens of metres of conventional accelerator can in principle be produced in centimetres of plasma, and it is this compactness that is moving the technology out of the physics laboratory and into commercial development.

Applications now being pursued include imaging and irradiation~\cite{albert2016}. The betatron radiation emitted by the bunch as it oscillates inside the plasma cavity provides a micron-scale, femtosecond hard X-ray source with a natural spatial coherence, which has been used for phase-contrast tomography of bone and of other low-contrast structures at a resolution that laboratory X-ray tubes cannot reach~\cite{cole2015}; the same capability underlies proposed non-destructive inspection of welds, additively manufactured components and battery cells. Scattering another high-intensity laser from the accelerated bunch yields a tunable, quasi-monoenergetic inverse-Compton source in the hundred-keV to MeV range, of interest for cargo screening and nuclear-material assay~\cite{powers2014}. In medicine, laser-driven very-high-energy electrons at $100$--$250\,\mathrm{MeV}$ penetrate deeply with a favourable depth-dose profile and can deposit a full treatment fraction in a single ultrashort pulse, which places them in the FLASH dose-rate regime~\cite{labate2020}. LPA bunches also reproduce in the laboratory the electron spectra used to qualify spacecraft electronics against single-event effects~\cite{hidding2011}. Compact LPA-driven photon sources are being commercialized by several ventures, including Tau Systems \cite{Hegelich2025NAS}, and free-electron lasing on an LPA beamline~\cite{wang2021,barber2025greater,kohrell2026over} points toward university- and industry-scale coherent X-ray sources.

\subsection{The Control Problem: Nonlinear Emission with Incomplete Diagnostics}
\label{sec:intro_control_problem}

Every one of these applications demands reproducibility at a given performance. A tomographic reconstruction is degraded by shot-to-shot variation in source flux and spectrum as surely as by insufficient flux; a radiotherapy dose must be delivered within a few per cent of prescription; screening and inspection are throughput problems that require hours of unattended operation at a fixed operating point. For this class of application the binding constraint is no longer gradient, guiding or injection, but the ability to hold an operating point across a shift and, when the point moves, to know why it moved. Stabilizing an LPA is the precondition for the technology leaving the laboratory.

Routine stability is hard for reasons specific to the laser-plasma interaction. A nonlinear map with measured inputs is a calibration problem; a linear map with hidden inputs is a regression problem; the laser-plasma accelerator is neither. The map from interaction-point conditions to beam observables is strongly nonlinear, so small upstream changes are not small downstream, and first-principles particle-in-cell simulation is far too expensive to run inside a real-time control loop. Parameters that determine beam quality are often very difficult for facility diagnostics to measure precisely. We organize many familiar examples by what is accessible in Table \ref{tab:params}.

\begin{table}[htb]
\centering
\caption{Parameters along an LPA beamline, grouped by what the control system knows about them.  Electron-beam transport is not separately instrumented and is omitted. The hidden column suggests targets for inference.  Environmental variables are measured but not (here) controlled, so act on the hidden column. The compressor and gas-jet rows carry no per-shot measurement because the drive laser is characterized upstream of the target chamber and interaction-point density measurements carry probe systematics. Of the commanded settings, only the backing pressure enters the emission map.}
\label{tab:params}
\small
\begin{tabular}{@{}>{\raggedright\arraybackslash}p{0.15\textwidth}
                  >{\raggedright\arraybackslash}p{0.20\textwidth}
                  >{\raggedright\arraybackslash}p{0.17\textwidth}
                  >{\raggedright\arraybackslash}p{0.38\textwidth}@{}}
\hline
Subsystem & Controlled (set-point) & Measured (per shot) & Hidden (neither set nor directly measured) \\
\hline
Drive laser &
Nominal focal spot $w_0$, compressor set-point duration $\tau_0$ &
Pulse energy $E_\ell$, spectrum, repetition rate, pointing (stabilized) &
Delivered amplitude $a_0$, amplifier thermal lens, mode quality $M^2$, wavefront error, fraction of energy in the focal spot, spatio-temporal coupling 
\\ \addlinespace[3pt]
Pulse compressor & 
Grating angles, compressor length &
--- & 
Residual chirp $\mathcal{C}$, grating thermal expansion, humidity-driven groove spacing, alignment drift, nonlinear phase ($B$-integral) 
\\ \addlinespace[3pt]
Laser beam transport & --- &  --- & 
Accumulated wavefront error, pointing and beam size at the focusing optic, focal plane position 
\\ \addlinespace[3pt]
Gas jet and plasma & 
Jet position (6D), backing pressure $p_{\mathrm{back}}$, blade coverage &  --- & 
Actual density profile $n_e(z)$ and its effective value for acceleration, shot-to-shot density fluctuation, shock position and width, gas temperature $T_{\mathrm{gas}}$, nozzle flow structure 
\\ \addlinespace[3pt]
Wakefield interaction & --- & --- & Injection phase, wake structure, beam loading 
\\ \addlinespace[3pt]
Electron-beam diagnostics & --- & $E$, $\Delta E/E$, $Q$; beam profile (recorded, unused) & Transverse and longitudinal phase space \\
\addlinespace[3pt]
Environment (facility) & --- & $T_{\ell}$, $T_{\mathrm{room}}$, RH & --- 
\\ \hline
\end{tabular}
\end{table}

To formalize this control challenge, we categorize the beamline variables into four operational classes:
\begin{enumerate}
    \item \emph{Commanded controls} $u_n$: dialable settings such as gas jet backing pressure $p_{\mathrm{back}}$, which shift the emission linearization point without introducing state uncertainty into the observation noise or its covariance;
    \item \emph{Measured beam observables} $y_n$: routine electron beam properties $y_n = [E, \Delta E/E, Q_b]$ (centroid energy, relative energy spread, and total charge);
    \item \emph{Hidden latent variables} $z_n$: effective interaction-point conditions $z_n = [a_0, n_{e18}, \mathcal{C}]$ (normalized laser field amplitude, electron density, and pulse chirp); and
    \item \emph{Unmeasured environmental drivers} $e_n$: slow ambient variations $e_n = [T_{\mathrm{laser}}, T_{\mathrm{room}}, \mathrm{RH}]$ (laser head temperature, room ambient temperature, and relative humidity).
\end{enumerate}
Because the drive laser is characterized prior to the target chamber and interaction-point density measurements may hide probe-delay systematics, $z_n$ remains unobserved and subject to drift driven by $e_n$. Start-up and tuning therefore rest on expert judgement and manual feedback that transfer poorly between facilities, and often between operating points at the same facility. The cost is paid in beam time: at a machine running a user programme, a drift that takes a morning to locate is a morning of delivered dose or delivered photons lost.

The response has been to bring data-driven methods to bear on the problem \cite{dopp2023}; two of the threads they identify are directly relevant here. The first is correlation analysis of operational records: Maier \emph{et al.}~\cite{maier2020} found correlations in the shot-to-shot energy variability of an LPA with measured properties of the drive laser, and stable operation over long durations has since been demonstrated at kilohertz repetition rate~\cite{rovige2020}. The second is automated optimization, principally Bayesian optimization, which has tuned accelerators at moderate dimensionality ($d=4$--$10$) over noisy objectives~\cite{shalloo2020,jalas2021,kirchen2021,duris2020}, including variants informed by reduced physical models~\cite{ferranpousa2023,hanuka2021}.

Bayesian optimization does well what it was designed to do. It maintains a surrogate, usually a Gaussian process, of an expensive and noisy objective as a function of control settings, and uses the surrogate's own uncertainty to choose the next setting to try, so that a good operating point can be reached in tens of shots rather than hundreds. Two features of that construction bound the information it can provide. Its coordinates are the control settings, not the physical state of the interaction, so every quantity the operator cannot dial, including the amplitude actually delivered at focus, the density actually present at the interaction point and the residual chirp on the compressed pulse, is absorbed into the noise term of the objective. Second, the objective is stationary, so the surrogate describes the machine as it was while the data were taken. Environmental drift violates both assumptions at once. The symptom is a loss of performance that re-optimization can recover, at the price of a new set-point and no statement of what moved: a warm chiller and a thermally drifting compressor are indistinguishable, because neither appears anywhere in the model. Performance degrades further when the control parameters are strongly correlated, or when the quantity of interest is a stability metric rather than a mean. Correlation analysis meets a matching limit from the other direction. The coefficients that account for centroid energy in one session change over subsequent sessions, which is itself evidence that variables outside the fit are moving~\cite{maier2020}. What is missing from both is not accuracy but attribution: neither answers the operational maintenance question of which subsystem to touch.

This paper takes the complementary position that a model useful for diagnosis must carry the physical state of the interaction explicitly. We therefore describe LPA operation with a \emph{latent state-space model}: a representation in which a small number of unobserved variables, collected in a latent state $z_n$, generate the measurements $y_n$ through an emission model, and evolve from shot to shot through a transition model driven by environmental inputs $e_n$.  Given the two models, a filter estimates $z_n$ recursively from the observations\footnote{The `filter' terminology for the method arose in signal processing, where the problem is separating the signal from unknown noise in a time series: the goal is an optimal sequential estimator or `filter' building upon the earlier frequency-domain Wiener filter where the relationship to basic high-pass or low-pass filters was more apparent.}. The construction is standard; what makes it diagnostic here is the choice of latent coordinates. We take $z_n$ to be the three effective interaction-point quantities $(a_0, n_{e18}, \mathcal{C})$, the normalized laser amplitude, the plasma electron density and the residual pulse chirp, which are the minimal set on which the leading-order blowout scalings depend, and infer them shot by shot from routine electron beam observations $y_n = (E, \Delta E/E, Q_b)$ using an extended Kalman filter. Because these coordinates are physical rather than nominal, naming the variable that moved indicates a subsystem. Keeping the emission and transition models structurally separate then splits diagnosis into two answerable questions: the Jacobian of the emission model identifies which latent variable moved, and comparison of competing transition models by predictive likelihood assigns that motion to a specific environmental mechanism. The output is not a better set-point but a statement of the form \emph{the energy fell because the delivered amplitude fell, and the amplitude fell because the laser head is running warm} from diagnostics facilities already record, and reported with an evidence margin that says how far it should be trusted.

This construction is not specific to the LPA. Its three ingredients, namely a small set of physical latent variables, an emission model mapping them to routine diagnostics records, and a family of competing transition models encoding known hardware couplings, can be assembled for any complex driver-target subsystem whose internal state is inferred rather than measured, and the same diagnostic chain should offer quantitative insights into performance degradation.

\subsection{Scope of This Work}
\label{sec:intro_scope}

We conduct a structural feasibility study to show that a small, physically parameterized state-space model, driven by routine diagnostics, suffices to isolate and attribute subsystem drift. The model is deliberately simplified and evaluated on synthetic data where ground truth is known. To test this structural concept, we introduce three core simplifications.
\begin{enumerate}
    \item \emph{Three-dimensional latent space}: The dozens of unmeasured parameters are collapsed onto a minimal three-dimensional latent state $z \in \mathbb{R}^3 = [a_0, n_{e18}, \mathcal{C}]$, representing the normalized laser amplitude, plasma electron density, and residual pulse chirp. The model thus identifies the effective physical quantity that moved rather than an unobservable upstream parameter;
    \item \emph{Two-stage physical emission model}: A reduced emission map $g(z, u)$ combining optical self-focusing dynamics with blowout acceleration scaling laws, with calibration constants aligned to typical performance~\cite{labun2025}; and
    \item \emph{Slaved and drift transition templates}: A set of candidate transition models encoding pre-specified hardware mechanisms, with physical coupling coefficients.
\end{enumerate}

The scope isolates the diagnostic problem.  We focus on fault detection, latent state localization, environmental attribution, and diagnostic resolution limits under measurement noise and driver magnitude.  Out-of-scope are active closed-loop feedback control and transverse phase space dynamics.  A demonstration on a simplified model with known ground truth is a necessary first step because the underlying argument is structural rather than quantitative. If the diagnostic chain cannot succeed when the data generator and inference filter share the same functional family and ground truth is known, it cannot be expected to function on operational machine data.

This formulation builds upon condition-monitoring and fault-diagnosis principles established across complex engineering systems~\cite{wang2017,bansal2025,amin2026}, adapting state-space isolation to laser-plasma beamlines. The remainder of the paper is organized as follows: Section~\ref{sec:model} constructs the physics model, detailing the latent state structure, emission map, and slaved transition dynamics (with the state-space schema presented in Section~\ref{sec:latent_state_structure}); Section~\ref{sec:chain} formulates the diagnostic pipeline; Section~\ref{sec:results} evaluates performance across synthetic benchmark scenarios; Section~\ref{sec:discussion} discusses structural limits and hazard cancellation; and Section~\ref{sec:limitations} details limitations and experimental validation pathways.

\section{Machine and physics models}
\label{sec:model}

\subsection{Latent State Structure}
\label{sec:latent_state_structure}

\begin{figure}[htbp]
\centering
\includegraphics[width=\textwidth]{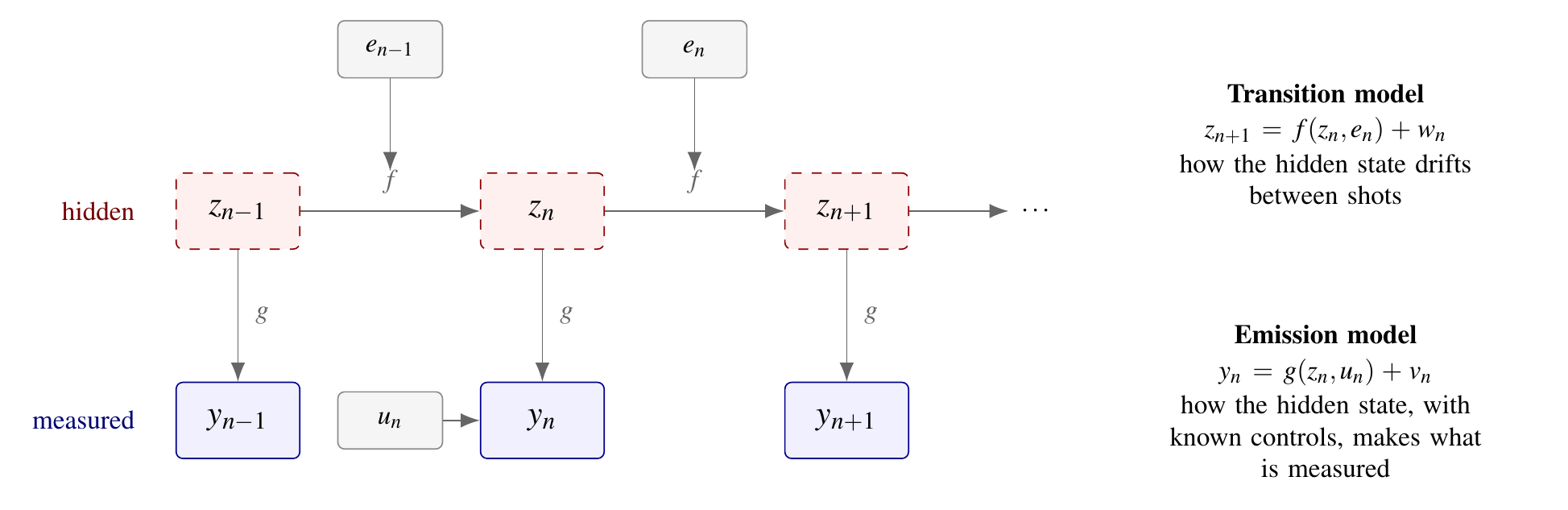}
\caption{The latent state-space structure. An unobserved state $z_n$ evolves under the transition model $f$, driven by environmental inputs $e_n$, and generates the measurements $y_n$ through the emission model $g$ together with the known controls $u_n$.}
\label{fig:ssm}
\end{figure}

Laser-plasma accelerator operation is formulated using a discrete-time latent state-space framework. The architecture, visualized in Fig. \ref{fig:ssm} separates the unobserved physical state at the interaction point from the routine electron beam measurements and environmental drivers.

The system dynamics are governed by two structurally decoupled models,
\begin{align}
y_n &= g(z_n, u_n) + v_n \quad &\text{(Emission Model)}, \label{eq:emission_gen} \\
z_{n+1} &= f(z_n, e_n) + w_n \quad &\text{(Transition Model)}, \label{eq:transition_gen}
\end{align}
where $n$ indexes individual shots. The vector $y_n = [E,\, \Delta E/E,\, Q]$ represents routine electron beam observations consisting of centroid energy $E$, relative energy spread $\Delta E/E$, and charge $Q$. The vector $u_n$ contains known control settings (such as gas jet backing pressure), while $e_n = [T_{\ell},\, T_{\mathrm{room}},\, \mathrm{RH}]$ records environmental parameters including laser-head temperature, room temperature, and relative humidity. The terms $v_n \sim \mathcal{N}(0, R)$ and $w_n \sim \mathcal{N}(0, Q)$ denote observation and process noise, respectively.

The latent state $z_n$ collapses the unmeasured beamline parameter space onto a three-dimensional vector defined at the interaction point $z_n =[ a_0, \tilde{n}_e, \mathcal{C}]$.  Here, $a_0 = |e \vec{E}_\ell| / (m_e c \omega_\ell)$ is the peak normalized laser vector potential at focus, measuring field intensity and relativistic drive strength.  The plasma electron density is normalized, $\tilde{n}_e = n_e / n_{\mathrm{ref}}$ relative to the reference density $n_{\mathrm{ref}} = 3.0 \times 10^{18}\,\mathrm{cm^{-3}}$.  Residual pulse spectral chirp, $\mathcal{C}$, is defined by its broadening effect on pulse duration $\tau_\ell = \tau_0 (1 + \mathcal{C}^2)$ relative to the transform-limited duration $\tau_0$.  These three effective quantities offer a minimal set parameterizing leading-order 3D blowout regime scalings. Complex upstream beamline variations, such as including laser mode quality ($M^2$), wavefront aberrations, thermal lensing in amplifier crystals, optical compressor grating expansion, and gas nozzle temperature shifts, are nonlinearly projected onto effective variations in $z_n$.

Keeping $g(z_n, u_n)$ structurally separate from $f(z_n, e_n)$ provides two key diagnostic benefits.  Since the emission model $g$ maps the interaction-point state directly to downstream observations $y_n$, its Jacobian $H = \partial g / \partial z$ enables shot-by-shot localization to identify which latent variable moved.  This decouples the detection of drift in the system from attribution.  Next, since the transition model $f$ governs how the latent state evolves under environmental forcing, comparing competing hardware-derived transition models by predictive likelihood identifies what mechanism caused the movement.

\subsection{Emission model}
\label{sec:emission_v1}

The static emission mapping $g(z, u)$ maps the latent interaction state $z = (a_0, n_{e18}, \mathcal{C})$ and control parameter $u = u_n$ to the physical observables $y = (E, \Delta E/E, Q_b)$.  Given its prevalence and relative simplicity, we start from the scaling for a matched, self-guided pulse in the blowout regime~\cite{lu2007}.  In this model, the dephasing-limited energy gain,
\begin{equation}
E \;\simeq\; \tfrac{2}{3}\,m_ec^2\,\frac{n_{\mathrm{cr}}}{n_e}\,a_0\,,
\label{eq:E_lu_base}
\end{equation}
contains the two relevant dependences: energy falls with density, because a denser plasma has a shorter dephasing length, and rises with laser amplitude. It is, however, a scaling for the \emph{matched in-plasma} amplitude, and it says nothing about energy spread, charge, or chirp.  We first construct a model for the in-plasma $a_0$.

Residual pulse chirp $\mathcal{C}$ is defined by its broadening effect on the temporal pulse duration $\tau_\ell = \tau_0 (1 + \mathcal{C}^2)$, reducing the vacuum field amplitude delivered to the plasma focus to
\begin{equation}
a_{\text{del}} = \frac{a_0}{\sqrt{1 + \mathcal{C}^2}} .
\label{eq:adel}
\end{equation}
The ratio of peak laser power $P$ to the critical power for relativistic self-focusing $P_{\text{cr}}$ defines the dimensionless power ratio $F \equiv P / P_{\text{cr}}$. Incorporating control parameter $u_n$ through the normalized density $\tilde{n} = \frac{n_{e18}}{3} u_n$ yields
\begin{equation}
F = \frac{a_0^2 K_0^2 \tilde{n}}{32 \left(1 + \mathcal{C}^2\right)} ,
\label{eq:power_ratio_F}
\end{equation}
where $K_0=7.01$ is a geometry factor, $P_{\rm cr} = 17.4\,(n_{\rm cr}/n_e)\ \mathrm{GW}$ is the critical power convention standard in the LPA literature, and $u_n = p_{\text{back}} / p_{\text{ref}}$ accounts for intentional backing pressure adjustments.

Under strong self-focusing the pulse contracts transversely. We obtain the map $a_{\rm lat}(F)$ from the paraxial eikonal solution of the relativistically self-focusing envelope equation~\cite{liu2019high}, $a=a_0(r,z)e^{iS}$, retaining only the relativistic term of the nonlinear refractive index and dropping the density-response correction, which presumes a weakly perturbed paraxial channel rather than the fully cavitated one considered here. (The variational reduction of the envelope equation agrees on the power-law exponent in $F$ but not on the numerical prefactor.) The self-trapping condition on the resulting beam-width equation gives
\begin{equation}
a_{\rm lat} = \sqrt{2\left[(4F)^{2/3}-1\right]} \;\xrightarrow[F\gg1]{}\; 2.25\,F^{1/3} ,
\label{eq:alat_eikonal}
\end{equation}
the same $F^{1/3}$ scaling from interpreting the matched-beam (blowout) condition $k_p w_{\rm eq} = 2\sqrt{a_{\rm lat}}$ together with power conservation $a_{\rm lat} w_{\rm eq} = a_{\rm del} w_0$ as a final equilibrium intensity,
\begin{equation}
a_{\text{lat}} = (8 F)^{1/3} = 2 F^{1/3} .
\label{eq:alat_matched}
\end{equation}
The two prefactors agree to within about $12\%$ at the reference operating point ($F\simeq5$). We adopt Eq.~\eqref{eq:alat_matched} as the working relation for simplicity.

Equation~\eqref{eq:alat_matched} is valid only over a bounded range of the latent state. Below $F\simeq1$ (i.e.\ $a_0\lesssim1$ at the nominal density) the pulse does not self-focus and the relation does not apply; well above the operating point $a_0\simeq1.8$, the single-mode Gaussian ansatz underlying both reductions above breaks down as the channel becomes more strongly evacuated than either solution assumes, so the map should not be extrapolated much beyond $a_0\sim 3$. The $(1+\mathcal{C}^2)^{-1}$ reduction of $F$ in Eq.~\eqref{eq:power_ratio_F} is similarly a leading-order account of chirp-induced pulse stretching and assumes $|\mathcal{C}|\ll 1$, so that higher-order spectral phase and pulse reshaping neglected do not matter.

Composing ($a_{\text{lat}} \propto F^{1/3}$) into Eq.~\eqref{eq:E_lu_base} with exponent $1/3$ yields the explicit functional dependence on latent state variables,
\begin{equation}
E = \mathcal{E}_0\,a_0^{2/3}\,\tilde{n}^{-2/3}\left(1 + \mathcal{C}^2\right)^{-1/3} ,
\qquad
\mathcal{E}_0 \;\equiv\; C_E\cdot\frac{2}{3}\,m_ec^2\,\frac{n_{\rm cr}}{n_{\rm ref}}\cdot 2\left(\frac{K_0^2}{32}\right)^{1/3} ,
\label{eq:E_v1_scaling}
\end{equation}
where $\mathcal{E}_0$ carries the units of energy.  Equation~\eqref{eq:E_v1_scaling} reproduces the matched scaling $E \propto P^{1/3} n_e^{-2/3}$. Requiring $E=680.4\ \mathrm{MeV}$ at the reference state $(a_0,\tilde n,\mathcal C)=(1.80,1,0)$ fixes $\mathcal{E}_0 = 459.8\ \mathrm{MeV}$.

The relative energy spread is modeled as an intrinsic injection floor combined with a gradient-controlled acceleration term:
\begin{equation}
\frac{\Delta E}{E} = 100\,\frac{\Delta E_{\min}}{E} + C_\sigma \frac{\sqrt{a_{\text{lat}}}}{1 + h(\mathcal{C},\tilde n)} ,
\label{eq:dE_v1}
\end{equation}
where $\Delta E_{\min} = 2\ \mathrm{MeV}$ sets an absolute injection-energy floor ($100\,\Delta E_{\min}/E$ in percent) and $C_\sigma=1.2\%$. This floor is the lower envelope $\Delta E_e \simeq \overline{eE_z}\,\Delta z_{\rm inj}$ of the injection-length argument identified against the measured performance envelopes of Ref.~\cite{labun2025}, found there to be essentially independent of plasma density; entering as $\Delta E_{\min}/E$, it is the only term in Eq.~\eqref{eq:dE_v1} that improves as the beam energy rises, consistent with a floor set at injection rather than one accumulated over acceleration. Downramp steepening and phase-space modification induced by residual chirp interacting with density are captured by
\begin{equation}
h(\mathcal{C},\tilde n) = \kappa_h \mathcal{C} \tilde{n}^{3/2} , \qquad (\kappa_h = 0.75) .
\label{eq:g_downramp}
\end{equation}
The mechanism and the density exponent follow the downramp-injection analysis of Ref \cite{pathak2012effect}: a positive (red-shifted-front) chirp lowers the local wake phase velocity, which acts as an effective steepening of the density downramp at fixed ramp geometry, and the resulting $\tilde n^{3/2}$ scaling follows from the density dependence of the plasma wavelength and etch-back length at fixed ramp profile. Other plasma-controlled parameters, shock position, injection phase, and downramp structure, remain outside the latent state $z$ and are not representable: $h$ captures only the beam response to chirp and density.  Independent drift in shock position or injection phase, from eg a shifting gas jet or a degrading nozzle, can still alias into $\kappa_h$ or be misattributed to whichever of $\mathcal C$ and $\tilde n$ best mimics its signature.

Trapped beam charge is described by a linear downramp injection scaling modified by chirp-induced downramp steepening,
\begin{equation}
Q_b = C_Q  \tilde{n}^{p_Q} \left(1 + \eta_Q h(\mathcal{C},\tilde n)\right) , \qquad (p_Q = 1, \; \eta_Q = 1),
\label{eq:Q_v1}
\end{equation}
with $C_Q=72.0$ pC chosen to reproduce the baseline operating point.  

\begin{figure}[htbp]
\centering
\includegraphics[width=\textwidth]{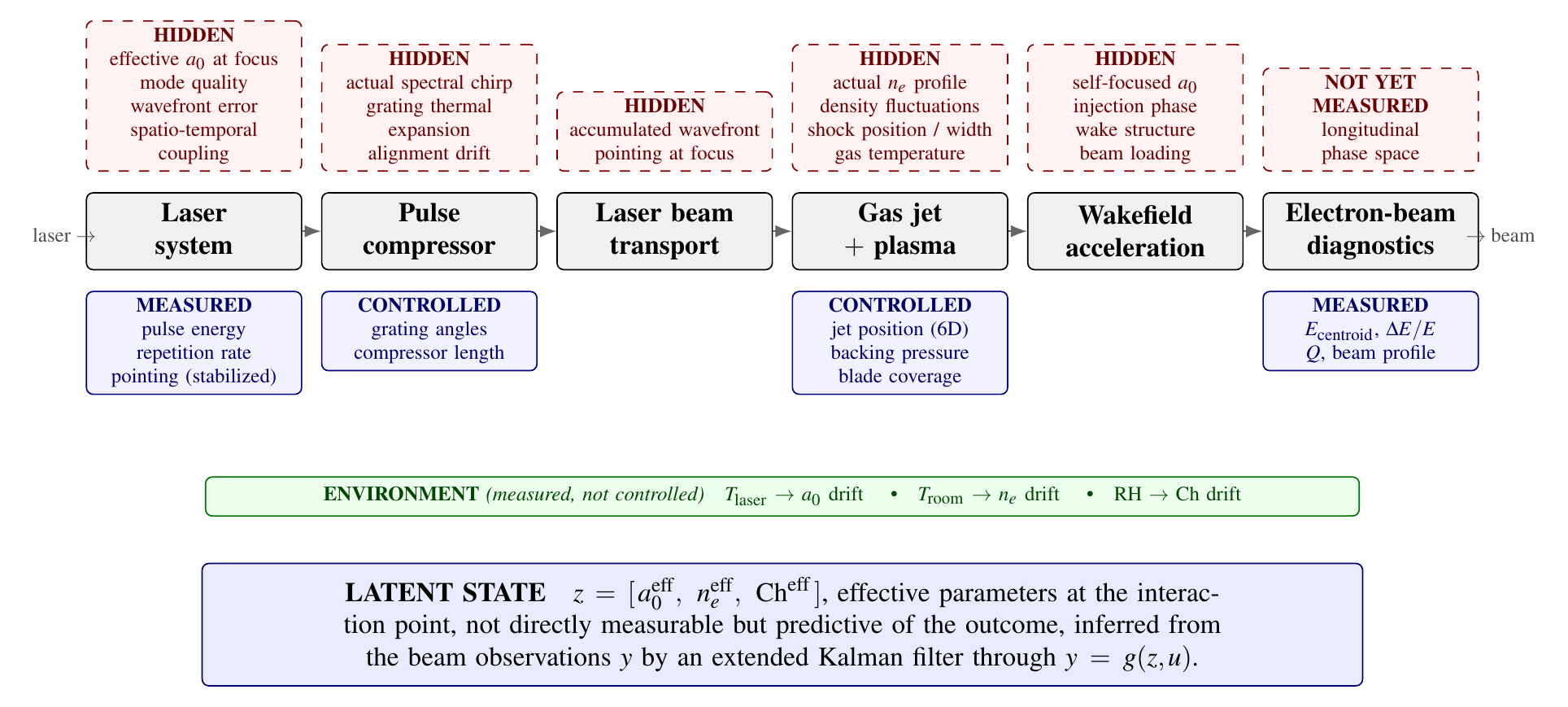}
\caption{An LPA beamline organized by what is knowable shot to shot. Above each subsystem (red, dashed) are hidden quantities, which are neither set nor diagnosed; below (blue) are controlled and measured ones. Environmental variables (green) can be measured but not controlled, and act on the hidden quantities rather than on the measured ones. Many physical parameters collapse into the three effective latent variables, as described in Table~\ref{tab:collapse}.}
\label{fig:beamline}
\end{figure}

The downramp-injection coupling $h(\mathcal{C},\tilde n)$ (Eq.~\ref{eq:g_downramp}) represents a first-order piece of the shock-position and injection-phase physics, so a drift in them can alias into the fitted coefficient $\kappa_h$, and hence into $\mathcal{C}$ or $\tilde n_e$, in addition to the calibration constants $C_\sigma$, $C_Q$.  A summary of how hidden quantities mentioned in text and in Table~\ref{tab:params} could impact the observables is organized visually in Fig. \ref{fig:beamline} and discussed with respect to model equations in Table \ref{tab:collapse}.  The list is a discussion and not expected to be conclusive.

\begin{table}[htbp]
\centering
\caption{How the hidden quantities of Table~\ref{tab:params} can reach observables.  The first eight rows are representable in the latent state, though the eighth enters as noise; the next three act on quantities outside $z$; the last is wholly unrepresented. The right-most column names the environmental driver and the corresponding competing model of Table~\ref{tab:models}. ``None logged'' marks a quantity that moves a latent coordinate with no measured driver, and therefore aliases into whichever candidate moves the same coordinate. 
}
\label{tab:collapse}
\footnotesize
\begin{tabular}{@{}>{\raggedright\arraybackslash}p{0.21\textwidth}
                  >{\raggedright\arraybackslash}c
                  >{\raggedright\arraybackslash}p{0.21\textwidth}
                  >{\raggedright\arraybackslash}p{0.26\textwidth}
                  >{\raggedright\arraybackslash}p{0.12\textwidth}@{}}
\toprule
Hidden quantity & Latent & Enters through & Signature & Driver / model \\
\midrule
Focal-spot quality: wavefront error, $M^2$, energy fraction, spatio-temporal coupling, pointing and beam size, focal plane position &
$a_0$ &
$a_0^{2/3}$ via $a_{\rm lat}(F)$, Eq. (\ref{eq:E_v1_scaling}); residual term in $\sqrt{a_{\rm lat}}/(1+h)$, Eq.~(\ref{eq:dE_v1}) &
$E$ falls; spread falls weakly; $Q_b$ unaffected &
none logged \\
\addlinespace[3pt]
Amplifier thermal lens &
$a_0$ &
as above &
as above, during warm-up &
$T_{\ell}$, $B_{aT}$, $M_1$ \\
\addlinespace[3pt]
Residual chirp (duration channel), some $B$-integral contributions &
$\mathcal{C}$ &
$(1+\mathcal{C}^2)^{-1/3}$ in $F\to a_{\rm lat}\to E$, Eqs.~(\ref{eq:power_ratio_F})-- (\ref{eq:E_v1_scaling}) &
second order and even in $\mathcal{C}$, so it vanishes at transform-limited operating point &
$\mathrm{RH}$, $B_{\Ch H}$, $M_2$ \\
\addlinespace[3pt]
Grating deformation: thermal expansion, humidity-driven groove spacing,
compressor alignment drift &
$\mathcal{C}$, $\tilde n_e$, $a_0$ &
$h=\kappa_h\mathcal{C}\,\tilde n_e^{3/2}$, Eq.~(\ref{eq:g_downramp}), entering $1/(1+h)$ in Eq.~(\ref{eq:dE_v1}) and $(1+\eta_Qh)$ in Eq.~(\ref{eq:Q_v1}); alignment channel via $\gamma_{ac}$, Eq.~(\ref{eq:couplings}) & 
first order and odd in $\mathcal{C}$, amplified by $\tilde n_e^{3/2}$: opposite-sign response in spread (falls) and charge (rises) for positive chirp.  
$\gamma_{ac}$ couples $\mathcal{C}\to a_0$, and hence weakly into $E$, through the transition model &
$\mathrm{RH}$, $B_{\Ch H}$, $M_2$ \\
\addlinespace[3pt]
Density at interaction point &
$\tilde n_e$ &
$\tilde n_e^{-2/3}$ via $a_{\rm lat}(F)$, Eq.~(\ref{eq:E_v1_scaling}); $\tilde n_e$, Eq.~(\ref{eq:Q_v1}); weak dependence through $\sqrt{a_{\rm lat}}$ and $\Delta E_{\min}/E$, Eq.~(\ref{eq:dE_v1}); modulates $h$, Eq.~(\ref{eq:g_downramp}) &
$E$ and $Q_b$ move oppositely, spread rises; density also scales any chirp-driven spread and charge response through $h$ &
$T_{\mathrm{room}}$, $B_{nT}$, $M_3$ \\
\addlinespace[3pt]
Gas temperature $T_{\mathrm{gas}}$ &
$\tilde n_e$ &
$\rho\propto1/T$ at fixed $p_{\mathrm{back}}$, so through $\tilde n_e$ as above &
indistinguishable from any other density excursion; $T_{\mathrm{room}}$ is the proxy &
$T_{\mathrm{room}}$, $B_{nT}$, $M_3$ \\
\addlinespace[3pt]
Shot-to-shot density fluctuation &
$\tilde n_e$ &
jitter $w_n$, Eqs.~(\ref{eq:Qw}), (\ref{eq:gain}) &
density signature above, but white and uncorrelated shot to shot, inflating innovations without biasing them & 
jitter, $Q_w$; no driver \\
\midrule
Shock position and width &
--- &
partially absorbed into $\kappa_h$, Eq.~(\ref{eq:g_downramp}); remainder into $C_\sigma$, $C_Q$ &
mimics a chirp- or density-modulated downramp response, or a residual density/chirp offset &
outside $z$ \\
\addlinespace[3pt]
Injection phase &
--- &
partially absorbed into $\kappa_h$, Eq.~(\ref{eq:g_downramp}); remainder into $C_\sigma$ &
spread and charge offset, read as a chirp- or density-correlated shift &
outside $z$ \\
\addlinespace[3pt]
Nozzle flow structure &
--- &
shape changes go into $C_Q$, $C_\sigma$ and the inferred $\tilde n_e$ level &
standing offset in charge and spread, degenerate with the emission calibration and absorbed by the profiled $b^\star$ of Sec.~\ref{sec:race} &
outside $z$ \\
\addlinespace[3pt]
Beam loading, wake structure, transverse and longitudinal phase space &
--- &
not represented &
correlated residuals in all three observables &
outside $z$ \\
\bottomrule
\end{tabular}
\end{table}

A couple of phenomenological features stand out from this emission model. Because $h(\mathcal{C}, \tilde{n})$ is odd in residual chirp $\mathcal{C}$, a positive chirp ($\mathcal{C} > 0$) steepens the density downramp, increasing trapped beam charge $Q_b$ via $(1 + \eta_Q g)$ while decreasing relative energy spread $\Delta E / E$ via $(1 + h)^{-1}$. Centroid energy $E$, by contrast, is even in chirp $\mathcal{C}$ within the present model, coupling to $\mathcal C$ only through the delivered power ratio $F \propto (1 + \mathcal{C}^2)^{-1}$.  The model carries no chirp-sign asymmetry in the dephasing-limited energy gain itself. Eq.~\eqref{eq:E_v1_scaling} omits a chirp-sign asymmetry in peak energy of order $10\%$ due to group-velocity and dephasing effects that is predicted by the same downramp analysis that motivates Eq.~\eqref{eq:g_downramp}. To the extent that this omitted asymmetry is small relative to the density- and amplitude-driven changes in $E$, the sign asymmetry visible in $Q_b$ and $\Delta E/E$ against the near-symmetry of $E$ remains a useful dependence for separating compressor-induced phase drift from laser-energy or gas-density variations; including the omitted dephasing term in a future revision would partially, though not entirely, erode this separation.

Evaluating the Jacobian matrix $H = \left.\frac{\partial g}{\partial z}\right\vert{}_{z_\star, u=1}$ at reference operating state $z_\star = (a_0^\star, n_{e18}^\star, \mathcal{C}^\star) = (1.80, 3.0, 0.0)$ and nominal control $u=1$ yields Table~\ref{tab:jacobian_v1}.  The fractional power dependence implies equal logarithmic weight, $\frac{\partial \ln E}{\partial \ln a_0} = +2/3$ vs. $\frac{\partial \ln E}{\partial \ln n_e} = -2/3$. Consequently, typical room temperature fluctuations affecting gas density ($T_{\mathrm{room}} \to n_e$) induce larger centroid energy shifts than laser thermal lens drift ($T_{\mathrm{laser}} \to a_0$).  Because the downramp coupling $h$ contains a term linear in $\mathcal{C}$, the partial derivatives $\frac{\partial (\Delta E / E)}{\partial \mathcal{C}} = -1.655\,\%/\text{rad}$ and $\frac{\partial Q_b}{\partial \mathcal{C}} = +54.0\,\text{pC/rad}$ remain non-zero at transform limit ($\mathcal{C} = 0$). This provides relative humidity drifts ($\mathrm{RH} \to \mathcal{C}$) first-order visibility in both energy spread and beam charge, enabling rapid filter tracking of pulse compressor alignment and grating temperature changes, subject to the aliasing caveat noted above.

\begin{table}[htbp]
\centering
\caption{Sensitivity matrix (Emission Jacobian) $H = \left.\frac{\partial g}{\partial z}\right\vert{}_{z_\star, u=1}$ evaluated at reference state $z_\star = (1.80, 3.0, 0.0)$ and control $u_n=1$.}
\label{tab:jacobian_v1}
\begin{tabular}{lrrr}
\hline
Observable $y_i$ & $\partial / \partial a_0$ & $\partial / \partial n_{e18}$ & $\partial / \partial \mathcal{C}$ \\
\hline
\textbf{Energy $E$ [MeV]} & $+252.0$ & $-151.2$ & $0.000$ \\
\textbf{Spread $\Delta E / E$ [\%]} & $+0.300$ & $+0.188$ & $-1.655$ \\
\textbf{Charge $Q_b$ [pC]} & $0.000$ & $+24.0$ & $+54.000$ \\
\hline
\end{tabular}
\end{table}

\subsection{Transition model}
\label{sec:transition}

The transition model describes how the hidden state moves between shots. Unlike the emission map, which is a fixed nonlinear function of the interaction-point state, the transition model represents a genuinely uncertain claim: which environmental pathway, if any, is pushing the latent state around. We formulate it as a small family of competing structural hypotheses, each supplying its own affine map from measured environmental excursions to a latent target and its own answer to a second, structurally separate question, namely whether the state has any memory of its own beyond what the environment dictates. The two questions turn out to have different correct answers, and keeping them distinct makes the resulting comparison interpretable.

\subsubsection{Slaved transition vs.\ mean reversion}
\label{sec:slaved}

The latent state at shot $n$ is the sum of an instantaneous, exogenous target and a persistent, unexplained offset, plus shot-to-shot jitter:
\begin{align}
z_n &= \mu_i(\Delta e_n) + \delta_n + w_n, & w_n &\sim \mathcal N(0,Q_w),
\label{eq:slave}\\[2pt]
\delta_n &= A_i\,\delta_{n-1} + w^{\delta}_n, & w^{\delta}_n &\sim \mathcal N(0,Q_\delta),
\label{eq:offset}
\end{align}
with $\Delta e_n \equiv e_n - e_\star$ the excursion from a chosen reference environment $e_\star=(T_{\rm laser}^\star,T_{\rm room}^\star,\mathrm{RH}^\star)=(35\,^\circ\mathrm{C},\,22\,^\circ\mathrm{C},\,45\,\%\mathrm{RH})$. The candidate mean is
\begin{equation}
\mu_i(\Delta e) = \big(\Ident + \gamma_{ac}\,\mathcal{E}_{13}\big)
\big(z_\star + G\,B_i\,\Delta e\big),
\qquad (\mathcal{E}_{13})_{jk} = \delta_{j1}\delta_{k3},
\label{eq:mu}
\end{equation}
with $z_\star$ the reference latent state, $B_i=\operatorname{diag}(\beta_{aT},\beta_{nT}, \beta_{\Ch H})$ the coupling matrix to be restricted to the channels active under a given hypothesis $M_i$, and $G=\operatorname{diag}(g_1,g_2,g_3)$ the gain matrix that will be profiled by the candidate race of Sec.~\ref{sec:race}.  Throughout the subscript $i\in\{0,1,2,3,...\}$ identifies the candidate hypothesis including the null, which is always $i\to 0$.  The internal term $\gamma_{ac}\mathcal{E}_{13}$ feeds the filtered chirp back into the amplitude once, after the environmental drive is applied.  Because $z_{\star,3}=0$ this is equivalent to the left-multiplication in \eqref{eq:mu} and introduces no cycle.

Here, $\gamma_{ac}$ is taken strictly as the {compressor alignment channel}: the same grating adjustment that leaves a residual chirp also tilts the wavefront and shifts the pointing, degrading the achieved $a_0$ through focal-spot quality rather than through pulse duration. The intensity loss due to pulse stretching is already explicit in the emission model through $1/\sqrt{1+\Ch^2}$ in $\adel$ and $1/(1+\Ch^2)$ in $F$. Since the two paths are mechanically independent, $\gamma_{ac}$ should ultimately be calibrated against wavefront or pointing data taken at known compressor positions rather than inherited, and its sign is not fixed a priori by the chirp physics itself.

Equations~\eqref{eq:slave}--\eqref{eq:mu} do not incorporate mean reversion in the model of the hardware. If $T_{\mathrm{laser}}$ has drifted from its believed set point, $a_0$ sits at its shifted value for as long as the temperature remains discrepant, and only an intervention moves it back; there is no restoring force internal to the laser that would pull it toward $z_\star$.  We assume the same holds component by component: the density is regulated toward a \emph{commanded} value (the known control input $u_n$) from which the latent becomes discrepant due to environmental effects; the chirp would revert only under compressor adjustment.  $\mu_i(\Delta e_n)$ is the state given the environment, and not the endpoint of some internal relaxation (eg an Ornstein–Uhlenbeck type model).  Following control theory, we describe this mechanism as \emph{slaving} and we confine memory to the additive offset $\delta_n$ in \eqref{eq:offset}.  Writing the drive as a recursion, $z_{n+1}=Az_n+Be_n+w_n$, would integrate the offset over shots without bound.

We define hypotheses $M_1$ through $M_4$ with $A_i=0$: the latent state has no memory beyond what the environment supplies at that shot. Three consequences follow. First, the transition Jacobian $F_n=\partial\delta_n/ \partial\delta_{n-1}=A_i$ vanishes, so the predicted offset covariance is $P_{n|n-1}=0$ and each shot will be analyzed as an independent nonlinear inversion of the emission map rather than propagated from the last. Second, because $z_n$ depends only on $e_n$ and $w_n$, the observations are conditionally independent given the environmental record, and the joint predictive density factorizes,
\begin{equation}
p\big(y_{0:N-1}\mid e_{0:N-1},M_i\big) = \prod_{n=0}^{N-1} p\big(y_n\mid e_n,M_i\big).
\label{eq:margfact}
\end{equation} 
Factorization ensures that we will obtain the exact log marginal likelihood of the session under $M_i$, up to the linearization of the emission model $g(z,u)$. Third, the offset profile used to guard against a miscalibrated diagnostic becomes exact rather than first order, since no residual feeds back into a later update.

The null candidate $M_0$ is the opposite choice: $A_0=\Ident$ and $B_0=0$, so the state is a pure random walk with no environmental drive. $M_0$ is not a special case of \eqref{eq:slave}--\eqref{eq:offset} but the same construction with the drive switched off, so it can be scored on the same basis as the physical candidates. By retaining memory in $M_0$ and not in the candidates, we gain the flexibility to track whatever the named mechanisms cannot explain.

\subsubsection{Separation of jitter and drift noise}
\label{sec:noise23}

Two physically distinct noise terms enter \eqref{eq:slave}--\eqref{eq:offset}, and conflating them can lead to incorrect interpretations. The process noise $w_n$ with variance
\begin{equation}
Q_w = \operatorname{diag}\big(\sigma_{\mathrm{proc}}^2\big),
\qquad
\sigma_{\mathrm{proc}} = (0.007,\;0.018,\;0.004),
\label{eq:Qw}
\end{equation}
represents shot-to-shot white jitter, redrawn independently every shot and does not persist. The offset noise $w_n^\delta$ with variance
\begin{equation}
Q_\delta = A_i\operatorname{diag}(\sigma_{\mathrm{drift}}^2)A_i\Tr,
\label{eq:Qdelta}
\end{equation}
represents the unmodelled, persistent random-walk component of drift that a candidate's structural mean $\mu_i$ does not capture.  In our cases, $A$ is diagonal 0s (non-null hypotheses) or 1s (null), so $Q_\delta=$diag($\sigma_{\rm drift}^2$) for the null hypothesis and zero otherwise.  The notation admits more general mixtures.  We set $\sigma_{\mathrm{drift}}= \sigma_{\mathrm{proc}}$ for simplicity, though they are independent.  We return in Sec.~\ref{sec:limitations} to what determining $\sigma_{\mathrm{drift}}$ would require.

Because the state noise $w_n$ does not accumulate, it does not belong in the predicted state covariance. It must still be accounted for, since a perturbation of $z$ is visible in the diagnostics.  Instead, it contributes to the predicted observation covariance, mapped through the emission Jacobian, where it increases the expected spread by exactly one shot's worth of jitter (see Eq.~\eqref{eq:gain} of Sec.~\ref{sec:filter} below). The consequence to note here is that the predicted spread remains a faithful statement of the candidate expectation, so a systematic mismatch between the new observations and that spread indicates the noise scales $\sigma_{\mathrm{proc}}$, $\sigma_{\mathrm{drift}}$ as poorly set. Refitting them is then a recognizable and separate repair.

\subsubsection{Candidate hypotheses and design tension}
\label{sec:candidates23}

Diagnosis proceeds by asking which transition structure best explains the observed record. We have five competing hypotheses to score against each other: the null $M_0$, in which the state is a pure random walk with no environmental drive; three single-mechanism models $M_1$, $M_2$ and $M_3$, each activating one environmental pathway; and the full causal model $M_4$, which activates all three plus the internal alignment coupling. Table~\ref{tab:models} summarizes them. All five share the emission model of Sec.~\ref{sec:emission_v1} and the noise covariances of Eqs.~\eqref{eq:Qw}--\eqref{eq:Qdelta}, and differ by exactly one structural assumption.  That structure assumption, namely choosing which components of $\delta$ retain memory and which columns of $B_i$ are switched on, makes the winner interpretable rather than merely best-fitting.

\begin{table}[htbp]
\centering
\caption{The five competing transition models. $M_0$ is the drift-only null against which
evidence is reported in Sec.~\ref{sec:race}; $M_1$--$M_4$ are slaving-only, with memory
switched off ($A_i=0$) so that any unmodelled persistence is left for the innovation
sequence to reveal rather than absorbed into $\delta$.}
\label{tab:models}
\small
\begin{tabular}{@{}llcccc c@{}}
\toprule
& & \multicolumn{3}{c}{diagonal of $B_i$} & & \\
\hline
Candidate & Mechanism & $\beta_{aT}$ & $\beta_{nT}$ & $\beta_{\Ch H}$ & $\gamma_{ac}$ & $A_i$\\
\hline
$M_0$ & random walk (null)                    & --- & --- & --- & --- & $\Ident$\\
$M_1$ & thermal, $T_{\mathrm{laser}}\to a_0$   & $\checkmark$ & --- & --- & --- & $0$\\
$M_2$ & humidity, $\mathrm{RH}\to \Ch$         & --- & --- & $\checkmark$ & --- & $0$\\
$M_3$ & gas, $T_{\mathrm{room}}\to n_{e18}$    & --- & $\checkmark$ & --- & --- & $0$\\
$M_4$ & full causal                           & $\checkmark$ & $\checkmark$ & $\checkmark$ & $\checkmark$ & $0$\\
\hline
\end{tabular}
\end{table}

The couplings should come from hardware physics rather than from a fit:
\begin{equation}
\beta_{aT} = -0.008\ ^\circ\mathrm{C}^{-1},\qquad
\beta_{nT} = -0.015\ ^\circ\mathrm{C}^{-1},\qquad
\beta_{\Ch H} = +3\times10^{-4}\ \%\mathrm{RH}^{-1},\qquad
\gamma_{ac} = -0.02,
\label{eq:couplings}
\end{equation}
representing respectively crystal heating and thermal-lens defocus ($T_{\mathrm{laser}}\!\to\!a_0$), gas density at fixed backing pressure scaling as $1/T$ ($T_{\mathrm{room}}\!\to\!n_{e18}$), grating groove-spacing response to humidity ($\mathrm{RH}\!\to\!\Ch$), and the compressor alignment channel discussed in Sec.~\ref{sec:slaved}.

Candidates $M_1$--$M_4$ are required to carry $A_i=0$. A candidate that is allowed both an environmental drive $B_i$ and free drift $A_i\neq0$ can absorb an unmodelled, persistent excursion into $\delta$. If every candidate in the hypothesis comparison were allowed to drift, we would not be able to define independent tests for the presence of unmodeled forcing (a screen) and the ability of one candidate to outperform the null or other hypotheses (a race). A mixed candidate, carrying both a nonzero $B_i$ and a nonzero $A_i$, is straightforward to construct but is excluded here for this reason.  We would have to derive a new statistical metric to score a mixed candidate alongside the rest.



\section{Diagnostic procedure and filter recursion}
\label{sec:chain}

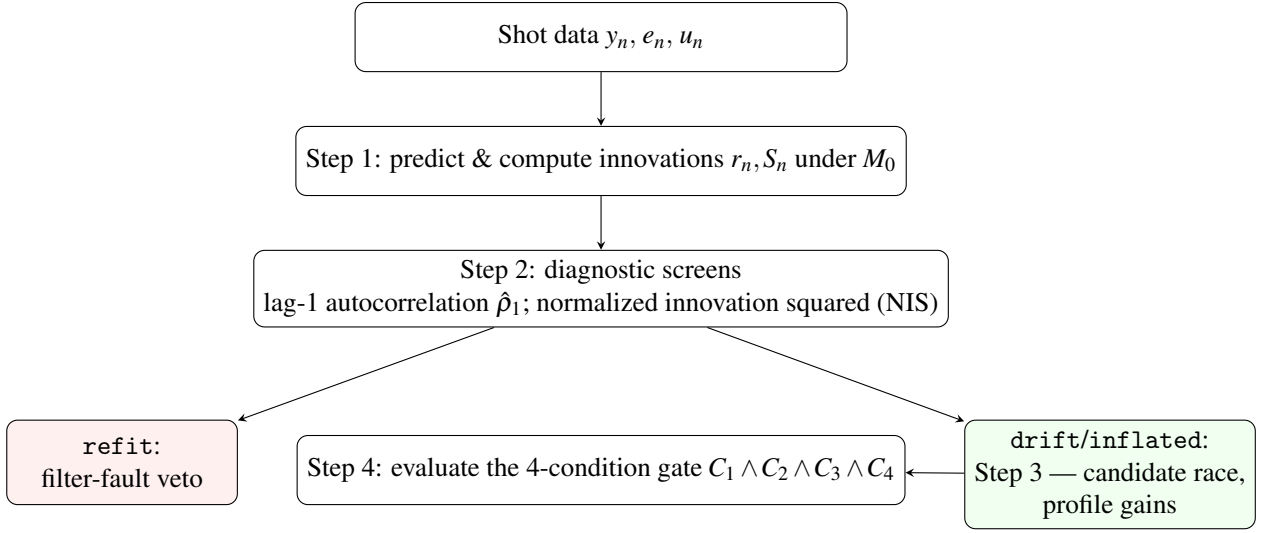
\begin{figure}[htbp]
\centering
\resizebox{\linewidth}{!}{%
\begin{tikzpicture}[
  node distance=7mm and 6mm,
  box/.style={draw, rounded corners, align=center, minimum width=6.4cm,
              minimum height=9mm, font=\small},
  veto/.style={draw, rounded corners, align=center, minimum width=3.0cm,
               minimum height=11mm, font=\small, fill=red!6},
  go/.style={draw, rounded corners, align=center, minimum width=3.4cm,
             minimum height=11mm, font=\small, fill=green!6},
  >=stealth
]
\node[box] (data) {Shot data $y_n,\,e_n,\,u_n$};
\node[box, below=of data] (predict)
  {Step 1: predict \& compute innovations $r_n,S_n$ under $M_0$};
\node[box, below=of predict]
  (screen) {Step 2: diagnostic screens\\ lag-1 autocorrelation $\hat\rho_1$; normalized innovation squared (NIS)};
\node[veto, below left=12mm and 2mm of screen] (veto) {\texttt{refit}:\\ filter-fault veto};
\node[go, below right=12mm and 2mm of screen] (race) {\texttt{drift}/\texttt{inflated}:\\ Step 3 --- candidate race,\\ profile gains};
\node[box, below=14mm of screen]
  (gate) {Step 4: evaluate the 4-condition gate $C_1\wedge C_2\wedge C_3\wedge C_4$};
\draw[->] (data) -- (predict);
\draw[->] (predict) -- (screen);
\draw[->] (screen) -- (veto);
\draw[->] (screen) -- (race);
\draw[->] (race) -- (gate);
\end{tikzpicture}}
\caption{The diagnostic chain. A \texttt{refit} verdict from screen~1 vetoes
the race and reports every downstream condition as \texttt{N/A}
(Sec.~\ref{sec:gate}); a \texttt{drift} verdict from either screen releases
it.}
\label{fig:chain}
\end{figure}

The diagnosis runs in four steps, shown in Fig.~\ref{fig:chain}: predict the state and score the information from the new shot (innovations) under the null transition $M_0$; screen those innovations for filter fault or drift; if the screen does not veto the record, race the competing transition models of Table~\ref{tab:models} against $M_0$ and profile their coupling gains; and evaluate a four-condition gate that turns a winning candidate into an addressable-drift declaration or withholds one.  We first define the quantities and calculations on which the inference is based.

\subsection{State prediction, update, and scoring}
\label{sec:filter}

The latent state is not measured, so every step needs the posterior probability density of the latent state at shot $n$, conditioned on the observations from shot $0$ through shot $n$ inclusive.  Because the emission model $g$ is nonlinear, the recursion closes only after linearizing about the predicted mean and propagating a Gaussian posterior: an extended Kalman filter (EKF)~\cite{sarkka2013}. Re-evaluating the linearization at every shot is the only approximation in the inference, and it ensures that the known control input shifts the working point rather than adding uncertainty. 

Each candidate $M_i$ of Table~\ref{tab:models} carries its own structural mean $\mu_i(\Delta e)$ and drift matrix $A_i$ (Sec.~\ref{sec:transition}) and runs its own filter on the offset $\delta_n\equiv z_n-\mu_i(\Delta e_n)$. 

\paragraph{Predict.} A subscript $n\mid n'$ reads ``at shot $n$, conditioned on data through shot $n'$,'' and only two cases occur: the \emph{prediction} $\hat\delta_{n|n-1}$ and $P_{n|n-1}$, formed before shot $n$ measurement arrives, and the \emph{filtered estimate} $\hat\delta_{n|n}$ and  $P_{n|n}$, formed after. The predicted mean $\hat\delta_{n|n-1}$ is what the candidate knows about $\delta_n$ before anything is observed, carrying the previous filtered estimate forward through whatever memory the candidate's drift matrix $A_i$ allows. The predicted covariance $P_{n|n-1}$ is the uncertainty attached to that prediction, i.e. how far $\delta_n$ can plausibly have wandered from $\hat\delta_{n|n-1}$.  The covariance can only grow between updates, because it transports the previous covariance $P_{n-1|n-1}$ through the same $A_i$ and adds the variance $Q_\delta$ due to one shot's drift.

The predicted \emph{state} then adds the candidate's structural mean $\mu_i(\Delta e_n)$, evaluated at that shot's own environmental record, on top of the predicted offset.  Thus, the prediction is the state implied jointly by the environment and the candidate's assumed persistence, before the shot's own beam diagnostics are consulted at all:
\begin{equation}
\hat\delta_{n|n-1}=A_i\,\hat\delta_{n-1|n-1},
\qquad
P_{n|n-1}=A_iP_{n-1|n-1}A_i^{\!\top}+Q_\delta,
\qquad
\hat z_{n|n-1}=\mu_i(\Delta e_n)+\hat\delta_{n|n-1}.
\label{eq:predict}
\end{equation}
Information enters only at the bar, so $P_{n|n}\preceq P_{n|n-1}$ and the gap between them is the new information brought by the shot.

\paragraph{Update.} The innovation $r_n$ compares the shot's outcomes $y_n$ against the model's prediction of them, mapped through the emission model at the predicted state.  The innovation is the only point at which data enters the filter, and it forms the basis of the downstream statistical model analysis. The innovration and the Jacobian $H_n$ are constructed on the observed channels $\mathcal{M}$ (all three by default) and re-evaluated at the predicted state and the shot's own control input
\begin{align}
r_n&=\Big[\,y_n-g\big(\hat z_{n|n-1},u_n\big)\Big]_{\mathcal{M}},
&
H_n&=\left[\frac{\partial g}{\partial z}\right]_{\hat z_{n|n-1},\,u_n,\,\mathcal{M}},
\label{eq:innov}\\[2pt]
S_n&=H_n\big(P_{n|n-1}+Q_w\big)H_n^{\!\top}+R,
&
K_n&=P_{n|n-1}H_n^{\!\top}S_n^{-1},
\label{eq:gain}\\[2pt]
\hat\delta_{n|n}&=\hat\delta_{n|n-1}+K_n\,r_n,
&
P_{n|n}&=\big(\Ident-K_nH_n\big)P_{n|n-1},
\label{eq:update}
\end{align}
initialized at $\hat\delta_{-1|-1}=0$ and $P_{-1|-1}$ nonzero only on the components $A_i$ allows to drift. Although not realized in our examples, a commanded pressure change would move the working point without modifying the covariance $P$. The innovation covariance $S_n$ is the predictive covariance of the observation, i.e. how much spread in $r_n$ the model expects from ignorance alone, before this shot's data arrives.  It is the sum of two contributions: state uncertainty carried forward from the previous shot and inflated by one shot of jitter, $P_{n|n-1}+Q_w$, mapped into observation units through $H_n$; and the observation noise $R$, which carries no information about the state at all, however large a given $r_n$ turns out to be. The gain $K_n$ uses that split to decide how much of $r_n$ to believe: $S_n^{-1}$ rescales the innovation by its predicted spread, and $P_{n|n-1}H_n^{\!\top}$ converts the rescaled quantity into a correction on $\delta_n$.  This way $K_n$ credits a given discrepancy to the state in proportion to how much of $S_n$ is state uncertainty rather than $R$. When $R$ dominates $S_n$, a large $r_n$ is more likely noise than signal and the gain damps it; when the state term dominates, the same $r_n$ is trusted more.

By structuring the hypotheses $M_1$--$M_4$ differently from the baseline $M_0$, we gain three advantages in the model comparison.
Since $A_i=0$ in the hypotheses, the covariance collapses to the noise $P_{n|n-1}=Q_\delta$ every shot, independent of the past. $K_n$ and $\hat\delta_{n|n}$ carry no information forward either.  Only the baseline $M_0$, with $A_0=\Ident$, keeps memory across shots.  The first advantage is that the relative log likelihood $\Delta\log L_i$ below becomes an exact log Bayes factor rather than a filtered proxy. Second, the observation-space offset profile of Sec.~\ref{sec:race} is exact rather than first order, since a residual that does not depend on past updates cannot feed back into itself. Finally it is cheap, since the covariance recursion short-circuits for four of the five candidates in the race.

\paragraph{Score.} After the update, each candidate has assigned a predictive density to the shot, providing the held-out likelihood as a by-product of the recursion. Summing the log of the held-out likelihood over the session and differencing against $M_0$ gives a running score each candidate can be ranked on
\begin{equation}
\ell_n^{(i)}=-\frac12\Big(r_n^{\!\top}S_n^{-1}r_n+\ln\det 2\pi S_n\Big),
\qquad
\Delta\log L_i=\sum_{n=0}^{N-1}\ell_n^{(i)}-\sum_{n=0}^{N-1}\ell_n^{(0)},
\qquad
\mathrm{BIC}_i=\Delta\log L_i-\tfrac12\,(\nu_i-\nu_0)\ln N,
\label{eq:score}
\end{equation}
with $\nu_0=0$ and $\nu_i$ the number of coupling gains profiled for candidate $i$.  BIC is short for Bayesian information criterion.  As shown in Sec.\ref{sec:race}, $\Delta\log L_i$ becomes the log Bayes factor of candidate $i$ against $M_0$ and a marginal likelihood over the whole session.  The offset $\delta_n$ is integrated over not maximized by the recursion, and so is not counted in $\nu_i$.  Counting it would penalize $M_0$ for the only physics it has, namely the bare (modeled) machine drift.

\subsection{Two-stage screening}
\label{sec:detect}

For a correctly specified model the innovations of $M_0$ are serially uncorrelated (i.e. white)~\cite{ljung1978}.  We check this two ways: one sensitive to the sign of an unmodelled trend and blind to its magnitude, the other sensitive to the magnitude and blind to the sign.  A \texttt{drift} verdict from either starts an attribution analysis; a \texttt{refit} verdict from the first vetoes the candidate detection, since a filter that cannot track its own null hypothesis cannot be trusted to compare structured alternatives against it.

\paragraph{Screen 1: lag-1 autocorrelation.} We monitor the autocorrelation
\begin{equation}
\hat\rho_1(k)=
\frac{\sum_{n\in W_k}\bigl(r_{n,E}-\bar r_k\bigr)\bigl(r_{n+1,E}-\bar r_k\bigr)}
     {\sum_{n\in W_k}\bigl(r_{n,E}-\bar r_k\bigr)^2},
\qquad
\bar r_k=\frac1W\sum_{n\in W_k}r_{n,E},
\label{eq:ac}
\end{equation}
here written for the energy channel of $r_n$, on non-overlapping windows $W_k$ of length $W=40$.  Overlapping windows with a first-crossing stopping rule give an unacceptably high session-level false-alarm rate even when the per-window exceedance rate is correctly calibrated, because the tests become heavily correlated. For a white series of length $W$, $\mathbb E[\hat\rho_1]=-1/(W-1)=-0.026$ rather than zero, and the two-sided band is centred there,
\begin{equation}
\mathcal B=\left[\,-\frac1{W-1}-\frac{z_\alpha}{\sqrt W},\;\;
                  -\frac1{W-1}+\frac{z_\alpha}{\sqrt W}\,\right],
\qquad z_\alpha=1.96,
\label{eq:band}
\end{equation}
and the two sides are routed to different conclusions, with a run-length requirement of $k$ consecutive exceedances:
\begin{equation}
\text{screen}_1=
\begin{cases}
\texttt{drift} & \hat\rho_1>\sup\mathcal B\ \text{for $k$ consecutive windows},\\[2pt]
\texttt{refit} & \hat\rho_1<\inf\mathcal B\ \text{for $k$ consecutive windows},\\[2pt]
\texttt{quiet} & \text{otherwise.}
\end{cases}
\label{eq:trigger}
\end{equation}
Recurring negative correlation signals a filter over-correcting shot to shot for jitter, which calls for updating the model of the noise.  Recurring {positive} correlation is the expected signature of drift.  Because the non-overlapping windows are independent, the session-level false-alarm rate per direction is $\sim(\alpha/2)^k$.  The parameter $k$ thus partially controls the minimum interval for detection as well as the tolerable false-alarm rate.

\paragraph{Screen 2: normalized innovation squared (NIS).} Lag-1 autocorrelation can be blinded by cancellation. For example, if the environment is driven faster than $M_0$'s response time, it cannot track the motion and instead lags behind it, producing a positive lag-1 correlation.  This positive autocorrelation can cancel the negative over-correction effect reducing the chance that the environmental driver is detected.  The second screen tests the magnitude of the innovation covariance instead,
\begin{equation}
\varepsilon_n=r_n^{\!\top}S_n^{-1}r_n\sim\chi^2_m,
\qquad m=\dim r_n,
\label{eq:nis}
\end{equation}
windowed as in \eqref{eq:trigger} and compared against the Wilson--Hilferty quantiles of $\chi^2_{mW}/(mW)$.  Higher-than-expected (\texttt{inflated}) innovation covariance could be unmodelled motion and hence routes to \texttt{drift} rather than to a second veto.  Lower-than-expected (\texttt{deflated} ) NIS is reported as \texttt{over-covariance}, i.e. a caution that the noise amplitudes $\sigma_{\mathrm{proc}}$ or $\sigma_{\mathrm{drift}}$ are set too high.

The combined verdict is summarized as
\begin{equation}
\text{screen}=
\begin{cases}
\texttt{refit} & \hat\rho_1\ \text{negative, sustained},\\[2pt]
\texttt{drift} & \hat\rho_1\ \text{positive, sustained, or NIS inflated},\\[2pt]
\texttt{over-covariance} & \text{NIS deflated},\\[2pt]
\texttt{quiet} & \text{otherwise.}
\end{cases}
\label{eq:screen}
\end{equation}
Testing this protocol under a correctly specified null ($2\times10^4$ trials, $400$-shot sessions, $\varepsilon_n\sim\chi^2_3$), we found the session-level false-alarm rate falls from $85.0\%$ for a simple screen 1-based first-crossing rule to $0.01\%$ for the two-screen test with $k=3$.  The false veto rate was $0.00\%$.

\subsection{Localization}
\label{sec:localize}

Localization relies on the filtered trajectory of the null candidate $M_0$.  The candidates have $A_i=0$, $K_n=0$ and $\hat\delta_{n|n}\equiv0$, so their predicted latent state $\hat z_n=\mu_i(\Delta e_n)$ is a deterministic restatement of the environmental record and carries no estimated excursion.  Writing $\hat z_n = z_\star+\hat\delta^{(0)}_{n|n}$ for $M_0$, we compute the spreads as the sample standard deviations over the record
\begin{equation}
\hat\sigma_{z_j}^2=\frac{1}{N-W}\sum_{n=W}^{N-1} \bigl(\hat z_{j,n}-\bar{\hat z}_j\bigr)^2 ,
\qquad
\bar{\hat z}_j=\frac{1}{N-W}\sum_{n=W}^{N-1}\hat z_{j,n},
\label{eq:sigmatraj}
\end{equation}
excluding an initial window to allow for transients in determining the filter covariance to decay.  We do not simply use the diagonal covariance $P_{jj}$, because that answers a different question, namely how well the filter knows $z_j$.  Eq.~\eqref{eq:sigmatraj} measures how far the prediction $\hat z_j$ actually moved.  We use the whole record after the first $W$ shots, not a window ending at a flagged shot.  The preliminary screen of Sec.~\ref{sec:detect} returns a verdict over non-overlapping windows and not a crossing shot, and on a \texttt{quiet} session there is no flagged shot to end at.  One nuance is that a windowed standard deviation understates a {monotone} excursion, for example returning $\Delta/\sqrt{12}$ for a linear ramp of difference $\Delta$.  Spreads of Eq.~\eqref{eq:sigmatraj} and peak displacements used to estimate recovery (below) therefore should not be compared directly.

Linearizing the observed $E$ about the window-mean state gives $\delta E\simeq\sum_j(\partial E/\partial z_j)\,\delta z_j$, and the share of the {observed} energy variance that first-order latent motion accounts for through $z_j$ is
\begin{equation}
\mathcal S_j=\frac{\bigl(\partial E/\partial z_j\bigr)^2\hat\sigma_{z_j}^2}{\widehat{\mathrm{Var}}(E)},
\qquad
\mathcal R=1-\sum_j\mathcal S_j ,
\label{eq:localize}
\end{equation}
with the Jacobian $H$ evaluated at $\bar{\hat z}$ and $\widehat{\mathrm{Var}}(E)$ the sample variance of the observed energy over the same window.  The shares are deliberately not normalized.  Against the observed variance, the remainder $\mathcal R$ is informative whether it comes out significantly high or low.  Positive $\mathcal R$ is observed scatter not explained by first-order latent motion, dominated by the observation noise $R$.  Negative $\mathcal R$ signals that the model's $Q_\delta$ may be larger than the true drift variance.  In other words, the model allows the state to wander in larger steps, which means the filter attributes a larger share of innovation to state motion rather than observation noise.  Thus when $\mathcal{R}<0$, the trajectory of Eq.~\eqref{eq:sigmatraj} is wider than the record it was inferred from.  Normalizing the shares would destroy this information. The absolute prediction of variance
\begin{equation}
\sigma_E^{\mathrm{model}}
=\Bigl[\textstyle\sum_j\bigl(\partial E/\partial z_j\bigr)^2\hat\sigma_{z_j}^2\Bigr]^{1/2}
\label{eq:sigmaEmodel}
\end{equation}
is directly comparable with the observed $\sigma_E$.  For $E\propto a_0^{\,p_a}\tilde n_e^{\,p_n}$ the ratio of shares is
\begin{equation}
\frac{\mathcal S_{a_0}}{\mathcal S_{\tilde n_e}}
=\left[\frac{p_a}{|p_n|}\,
\frac{\hat\sigma_{a_0}/a_0^\star}{\hat\sigma_{\tilde n_e}/\tilde n_e^\star}\right]^{2},
\qquad p_a=+\tfrac23,\quad p_n=-\tfrac23,
\label{eq:shareratio}
\end{equation}
independent of the calibration constants. The exponents being equal in magnitude in this particular emission model will have consequences for localization in test runs below. 

\subsection{Candidate race and family margin}
\label{sec:race}

A \texttt{drift} or \texttt{over-covariance} verdict initiate the attribution analysis: each candidate in Table~\ref{tab:models} is filtered by \eqref{eq:predict}--\eqref{eq:update} and scored by \eqref{eq:score}. The basis for summing one-step-ahead terms is the prediction-error decomposition, which is exact,
\begin{equation}
p\bigl(y_{0:N-1}\mid M\bigr)=\prod_{n=0}^{N-1}p\bigl(y_n\mid y_{0:n-1},M\bigr),
\end{equation}
so $\Delta\log L_i$ is a log Bayes factor of candidate $i$ against $M_0$.  For slaving-only candidates, it is equally the log marginal likelihood: with $A_i=0$, $\delta$ carries no memory, so the $y_n$ are conditionally independent given the environment record and $p(y_{0:N-1}\mid e_{0:N-1},M_i)=\prod_np(y_n\mid e_n,M_i)$ holds identically. Nothing is lost to the filtering approximation there beyond the linearization of the Jacobian $H_n$ itself.

Note that scoring each candidate at its literal hardware coupling $\beta_{aT},\beta_{nT},\beta_{\Ch H}$ would test whether that coefficient is precise rather than whether the mechanism is present at all.  To guard against this failure-by-over-specification, we profile one multiplicative gain $g_i$ per mechanism, $B_i\to G_iB_i$ by coordinate ascent on $\Delta\log L_i$.

The candidate models can differ by mechanisms whose $1\sigma$ effect on the observable is smaller than the observation noise, so exact-model identification is neither achievable nor attempted.  The subsystem is reported, through the margin of the family $\mathcal F$ containing the winning candidate over the best candidate outside it,
\begin{equation}
\Delta_{\mathcal F}=\max_{m\in\mathcal F}\mathrm{BIC}_m
                    -\max_{m\notin\mathcal F}\mathrm{BIC}_m .
\label{eq:family}
\end{equation}

\subsection{Limits}
\label{sec:limits}

Suppose candidate $M_i$ captures a deterministic component $\delta_n$ of the predicted energy that other candidates miss.  The cross term vanishes in expectation and Eq.~\eqref{eq:score} gives
\begin{equation}
\mathbb E\bigl[\Delta\log L_i\bigr]\simeq
\underbrace{\frac{N}{2}\,\ln\frac{\det S^{(0)}}{\det S^{(i)}}}_{\text{structural}}
\;+\;\frac12\sum_{n=0}^{N-1}\frac{\bigl(\delta^{(0)}_n\bigr)^2-\bigl(\delta^{(i)}_n\bigr)^2}
{S_{EE}} .
\label{eq:evidence}
\end{equation}
relative to the null.  The first term should not be interpreted as evidence for a mechanism.  It arises from $M_0$ carrying the converged predicted covariance $P_{n|n-1}$, whereas the candidates $M_i$ do not.  In a 400-shot test session below, it contributes $0.344$ nats per shot to every candidate whether or not it explains anything (\emph{nat} is a logarithmic unit of information defined using the natural logarithm, representing the amount of information gained from an event occurring with probability 1/$e$).  Read against the null, $\Delta\log L_i$ is therefore not on a mechanism-evidence scale.

Mechanism evidence is supplied by the family margin $\Delta F$.  Both sides of Eq.~\eqref{eq:family} are slaving candidates sharing $P_{n|n-1}=0$ and, to the accuracy of the shared linearization, a common $S$, so the structural term cancels identically and
\begin{equation}
\mathbb E\bigl[\Delta_{\mathcal F}\bigr]\simeq
\frac{N}{2\,S_{EE}}\Bigl(s_{\mathcal F}^{2}-s_{\neg\mathcal F}^{2}\Bigr),
\qquad S_{EE}=40.8\ \mathrm{MeV^2},
\label{eq:margin_scale}
\end{equation}
with the rms energy signature $s_M$ for a mechanism $M$ or family $\mathcal{F}$ defined
\begin{align}\label{eq:signature}
s_M= \bigl|\frac{\partial E}{\partial z_j}g_kB_{i,jk}\bigr|\,\mathrm{rms}(\Delta e_k), \quad 
s_{\mathcal{F}}=\bigl(\sum_{M\in \mathcal{F}}s_M^2\bigr)^{\!1/2}.
\end{align}
Evidence grows linearly in shot count and quadratically in the size of the signal, but only in the difference of the two signatures.  A mechanism competing against a background of comparable size accumulates nothing no matter how long the session runs.

\subsection{Four-condition addressable-drift gate}
\label{sec:gate}

An addressable drift is declared, and an intervention recommended, if and only if
\begin{equation}
\underbrace{\text{no filter fault, race invoked}}_{C_1}
\;\wedge\;
\underbrace{\Delta_{\mathcal F}>10}_{C_2}
\;\wedge\;
\underbrace{\delta E_{\mathcal F}>\mathrm{TOL}_E}_{C_3}
\;\wedge\;
\underbrace{\mathcal F\in\text{ACTIONS}}_{C_4},
\label{eq:gate}
\end{equation}
where $\delta E_{\mathcal F}$ is the peak-to-peak swing in centroid energy that the winning candidate's deterministic mean predicts across the session,
\begin{equation}
\delta E_{\mathcal F} =\max_{n}\,E\bigl(\mu_{i^\star}(\Delta e_n)\bigr) -\min_{n}\,E\bigl(\mu_{i^\star}(\Delta e_n)\bigr),
\qquad
i^\star=\operatorname*{arg\,max}_{i\in\mathcal F}\mathrm{BIC}_i ,
\label{eq:peakswing}
\end{equation}
evaluated at the profiled gains $G$ and through the full nonlinear emission model rather than its linearization.  Channel contributions decompose $\delta E\simeq\sum_{jk}(\partial E/\partial z_j)\,g_k B_{i,jk}\, (\Delta e_k^{\max}-\Delta e_k^{\min})$, with a further term $(\partial E/\partial a_0)\,\gamma_{ac}\,g_3\beta_{\Ch H}\,\Delta\mathrm{RH}$ for $M_4$, and the decomposition identifies the subsystem.

Three features of this protocol:  First, any persistent offset of the operating environment from the reference is degenerate with the emission calibration and is removed by the profiled $b^\star$ of Sec.~\ref{sec:race}.  Re-introducing it here would confound the score.  Second, the conclusion assumes the winning family is wholly responsible for the observed discrepancy, which becomes an upper bound on what an intervention recovers.  The estimated recovery is
\begin{align}
\delta E_{jk}&=\frac{\partial E}{\partial z_j}\,g_kB_{i,jk}\,\Delta\bar e_k , \qquad 
\Delta\bar e&=\frac1W\sum_{n=N-W}^{N-1}e_n-e_\star ,
\label{eq:recovery}
\end{align}
The recommendation is falsifiable: the operator intervenes, the energy is measured again, and the prediction is met or it is not.

Finally, the two roles of the screen in $C_1$ are not symmetric. A \texttt{refit} verdict vetoes, because it implies that the shared noise model is wrong, every candidate in the race inherits it, and the whole comparison is scored under a misspecified innovation covariance.  For this verdict, we should report $C_2$ through $C_4$ as \texttt{N/A}. A \texttt{quiet} verdict, by contrast, is not a veto.  In at least one synthetic session, the screen returned quiet on autocorrelation while the attribution analysis discovered the correct driver with high significance and gain profiling identified the correct value. The screen thus decides whether the comparison is worth running under a compute budget, not whether the race's answer counts once it has run.

\section{Results on synthetic sessions}
\label{sec:results}

No measured data enters this paper. Synthetic sessions are used because scoring a diagnosis requires ground truth about which mechanism was active, which operational records do not supply. The generator is stated here in full, so that the overlap between it and the inference model, which Sec.~\ref{sec:limitations} identifies as the principal caveat on everything that follows, can be inspected rather than taken on trust.  The selected dimensionful numbers are thought be to reasonable choices, and not representing data from any particular facility.

The environmental drivers reproduce behaviour seen in operations: a warm-up transient in the laser-head temperature, and slow oscillations in room temperature and relative humidity,
\begin{align}
T_{\ell}(n)&=
\begin{cases}
35+6\bigl(1-e^{-n/18}\bigr)\,^{\circ}\mathrm{C}, & n<60,\\[2pt]
41+A_T\sin(2\pi n/P)+0.12\,\xi_n\ ^{\circ}\mathrm{C}, & n\ge60,
\end{cases}
\label{eq:genT} \\[3pt]
T_{\mathrm{room}}(n)&=22+1.8\sin(2\pi n/200)+0.08\,\xi'_n\ ^{\circ}\mathrm{C},
\qquad
\mathrm{RH}(n)=45+4\sin(2\pi n/300+0.8)+0.2\,\xi''_n\ \%,
\label{eq:genE}
\end{align}
with $\xi,\xi',\xi''$ independent standard normals. The latent state follows the couplings of Eq.~(\ref{eq:couplings}) contemporaneously, with an optional chirp ramp representing optic degradation,
\begin{align}
C_{\tau,n}&=B_{CH}\,\Delta\mathrm{RH}_n+\rho\,\max(0,\,n-n_{\mathrm{ramp}})+w_{C,n},
\label{eq:genC}\\[2pt]
a_{0,n}&=1.80+B_{aT}\,\Delta T_{\ell,n}+\gamma_{aC}\,C_{\tau,n}+w_{a,n},
\qquad
\tilde n_{e,n}=1.00+B_{nT}\,\Delta T_{\mathrm{room},n}+w_{n,n},
\label{eq:genz}
\end{align}
and the observations are $y_n=g(z_n)+v_n$ with process and observation noise
\begin{equation}
w\sim\mathcal N\bigl(0,\;\mathrm{diag}(0.007,\,0.006,\,0.004)^2\bigr),
\qquad
v\sim\mathcal N\bigl(0,\;\mathrm{diag}(5.5\,\mathrm{MeV},\,0.22\%,\,2.2\,\mathrm{pC})^2\bigr),
\label{eq:gennoise}
\end{equation}
the densities expressed in the normalized variable $\tilde n_e$ throughout. Individual mechanisms are switched off by zeroing the corresponding coupling, e.g. generating the single-mechanism sessions of Sec.~\ref{sec:specificity}. The chirp ramp is active only in Secs.~\ref{sec:session} and~\ref{sec:excitation}, with the onset and rate stated in each.

After the warm-up the laser head does not return to the reference environment $e_\star=(35\,^{\circ}\mathrm{C},\,22\,^{\circ}\mathrm{C},\,45\%)$ at which the emission model was calibrated; it settles at $41\,^{\circ}$C. Both sessions of Sec.~\ref{sec:session} therefore run under a permanent $+6\,^{\circ}$C offset in $T_\ell$, with a modulation of amplitude $A_T$ superposed on it. Under a transition model that predicts the latent level absolutely, rather than its increments, it enters every candidate's residual as a constant, and a constant residual in observation space is degenerate with the emission calibration constants $C_E$, $C_\sigma$, $C_Q$. Scoring it as evidence about mechanism would therefore be an error of the same kind as scoring the calibration. All scores reported below are consequently profiled over a constant observation-space offset, and Sec.~\ref{sec:session} reports what happens when they are not.

\subsection{Two basic sessions}
\label{sec:session}

Here we study two sessions  differing only in the post-warm-up excursion of Eq.~(\ref{eq:genT}):
\begin{center}
\small
\begin{tabular}{@{}llll@{}}
\toprule
& $A_T$ & $P$ [shots] & character \\
\midrule
reference & $0.9\,^{\circ}$C & $140$ & ambient; \\
dithered  & $6.0\,^{\circ}$C & $30$  & $T_\ell$ deliberately excited\\
\bottomrule
\end{tabular}
\end{center}
Both run $400$ shots at seed $42$ with a chirp ramp of $6\times10^{-4}$ per shot from shot $250$. The pair is chosen so that the two possible outcomes of the gate are both exhibited on a session whose ground truth is known, and so that the difference between them is attributable to one generator argument.

\begin{figure}[htbp]
\centering
\includegraphics[width=\textwidth]{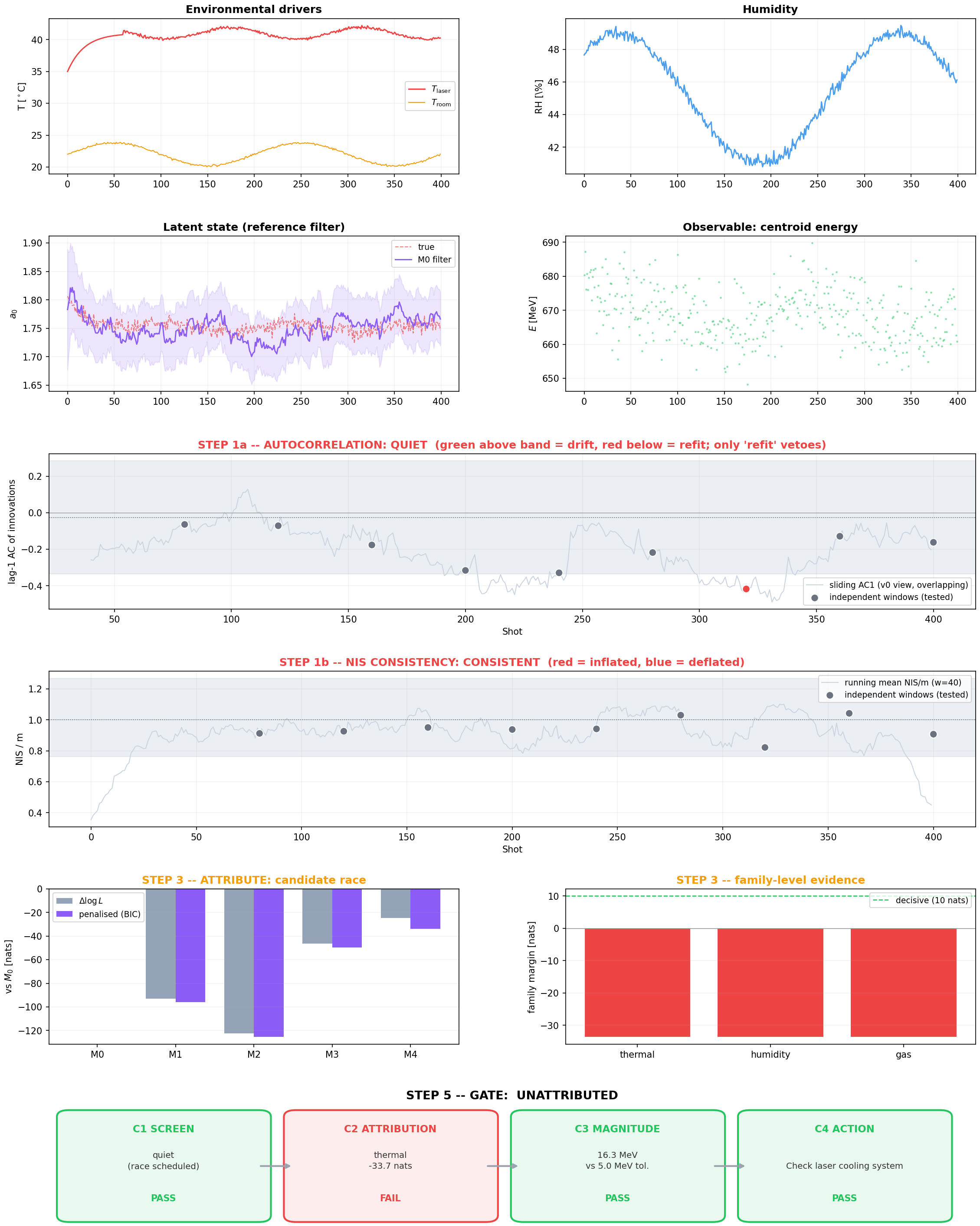}
\caption{The reference session ($A_T=0.9\,^{\circ}$C, $P=140$, $400$ shots, seed $42$), the full chain. \emph{Rows 1--2:} environmental drivers, the humidity channel, the $M_0$ filter tracking the hidden $a_0$, and the observed centroid energy. \emph{Rows 3--4:} the two screens, both quiet.  The tested non-overlapping windows lie inside their bands except for a single autocorrelation window at $-0.42$. The grey curve is the sliding statistic calculated from overlapping windows, shown for comparison. \emph{Row 5:} the candidate race and family margins. Every alternative loses to the null. \emph{Row 6:} the gate. C2 fails at $-33.7$ nats and the verdict is \textsc{unattributed}; no work order issued.}
\label{fig:session}
\end{figure}

\begin{figure}[htbp]
\centering
\includegraphics[width=\textwidth]{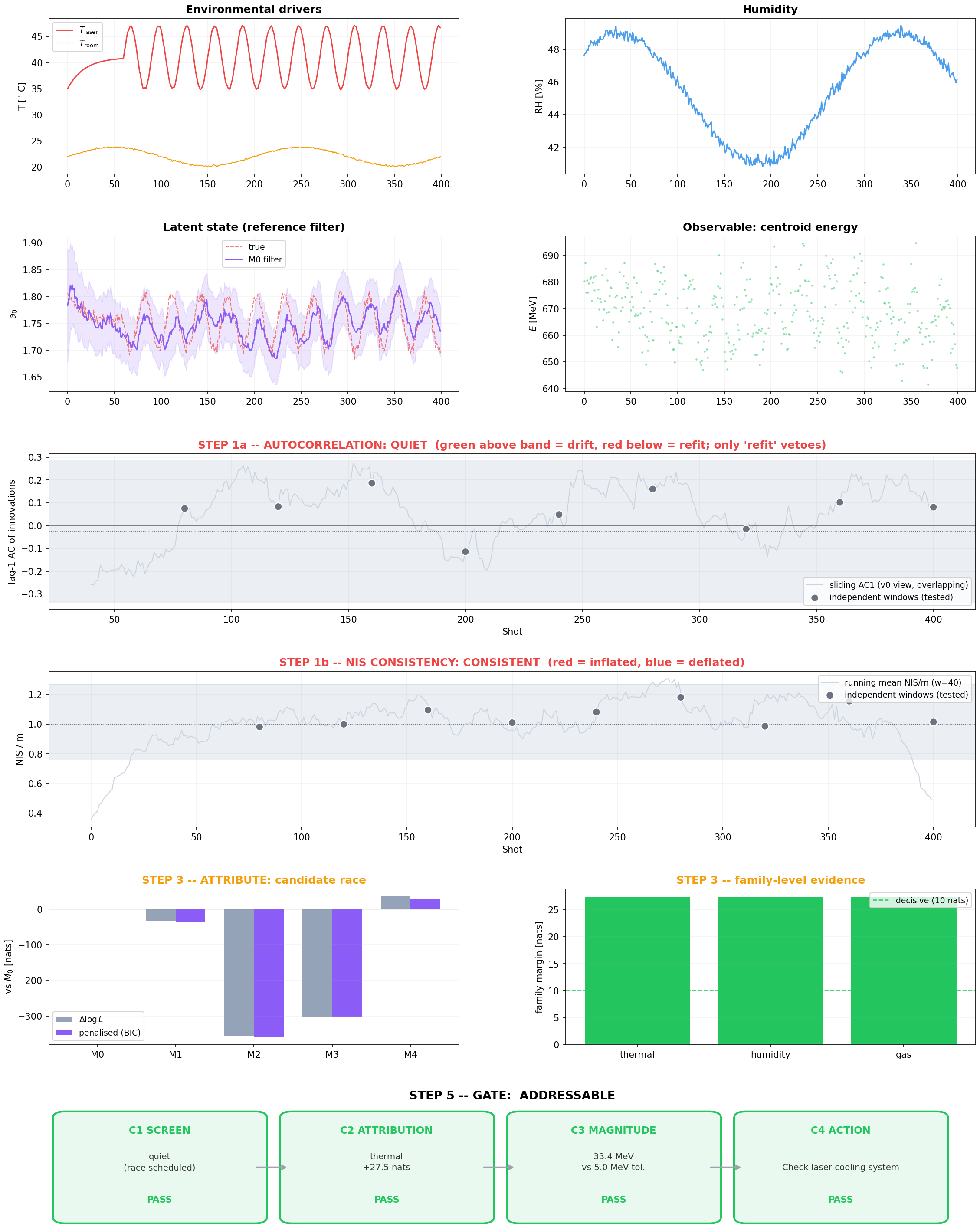}
\caption{The dithered session ($A_T=6.0\,^{\circ}$C, $P=30$; otherwise identical to Fig.~\ref{fig:session}). The screens remain quiet: the random-walk filter absorbs the excitation while the race resolves it. $M_4$ clears the null by $+27.5$ nats after penalty and the gate passes all four conditions. The family-margin panel comes out equal on all familes: the tie is broken by enumeration order, not by evidence.}
\label{fig:dither}
\end{figure}

Figures~\ref{fig:session} and~\ref{fig:dither} show the full chain on the reference and dithered sessions respectively. The two sessions reach opposite verdicts: the reference session is declared \textsc{unattributed} and issues no work order; the dithered session is declared \textsc{addressable} at a family margin of $+27.5$ nats. The only difference is $5.1\,^{\circ}$C of deliberate excitation.

Comparing $20$-shot means, the centroid energy falls by $5.7\,\mathrm{MeV}$ across the warm-up transient and by $10.3\,\mathrm{MeV}$ over the reference session as a whole.  The dithered session gives $10.4$ and $11.6\,\mathrm{MeV}$. Single-shot endpoints differ from these by up to $15\,\mathrm{MeV}$, which is why a smoothed estimator is quoted. The observed energy variance over the analysis window is $51.9\,\mathrm{MeV^2}$ on the reference session and $117.8\,\mathrm{MeV^2}$ on the dithered one, a factor of $2.3$ arising from the dither.

\paragraph{Detect.}
Neither session fires the screen. On the reference session the nine non-overlapping windows of Eq.~(\ref{eq:ac}) return lag-1 autocorrelations $(-0.06,-0.07,-0.18,-0.32,-0.33,-0.22,-0.42,-0.13,-0.16)$ against a band $[-0.336,+0.284]$ centred on the small-sample mean $-1/(W-1)=-0.026$ for $W=40$. The dithered session returns $(+0.08,+0.08,+0.19,-0.11,+0.05,+0.16,-0.02,+0.10,+0.08)$, entirely inside the band. The normalized innovation squared is consistent on both, running $0.82$ to $1.04$ and $0.98$ to $1.18$ against a band $[0.763,1.268]$.


\paragraph{Localize.}
On the reference session the filtered state averaged over the analysis window is $\bar z=(1.748,\,0.998,\,+0.020)$, at which the energy row of the Jacobian is
\begin{equation}
\frac{\partial E}{\partial a_0}=+254.8,
\qquad
\frac{\partial E}{\partial \tilde n_e}=-446.6,
\qquad
\frac{\partial E}{\partial \mathcal{C}}=-8.7\ \ \mathrm{MeV},
\label{eq:jacsession}
\end{equation}
the last being small and of a sign set only by the slightly positive filtered chirp; it vanishes identically at $\mathcal{C}=0$, so neither its size nor its sign should be read as physical. 

Note the reference session's shares sum to $133.8\%$. 
First-order latent motion cannot account for more variance than was observed, and a negative remainder is a diagnostic on the filter and not a property of the session.  The over-wide $\hat\sigma_{\tilde n_e}$ is being counted against a smaller observed variance. The dithered session, in which real excitation dominates the observed variance, gives a remainder of $+13.7\%$ and is unremarkable. We conclude that the offset random-walk scale may be set too loose relative to the shot-to-shot jitter.  This is an example of how the analysis framework will need to be tuned to each machine.  Until it is set deliberately, localization shares should be read as ordering the channels and not as a variance budget with a strict sum rule.  The two observables localize to different variables by construction.  


\paragraph{Attribute.}
The attribution analysis race is run on both sessions, over the full observation vector, with the per-mechanism coupling gains profiled and the scores penalized by $\tfrac12\,\Delta\mathrm{dof}\,\ln N$ with $\ln 400=5.99$:

\begin{center}
\small
\begin{tabular}{@{}lrrrrrr@{}}
\toprule
& \multicolumn{3}{c}{reference} & \multicolumn{3}{c}{dithered} \\
\cmidrule(lr){2-4}\cmidrule(lr){5-7}
candidate & gain $(a_T,n_T,C_H)$ & $\Delta\log L$ & BIC & gain & $\Delta\log L$ & BIC \\
\midrule
$M_0$ random walk        & ---                 & $0.00$    & $0.00$    & ---                 & $0.00$    & $0.00$ \\
$M_1$ thermal            & $(1.3,\,1,\,1)$     & $-93.14$  & $-96.13$  & $(1.0,\,1,\,1)$     & $-32.93$  & $-35.93$ \\
$M_2$ humidity           & $(1,\,1,\,2.5^\ast)$& $-122.52$ & $-125.51$ & $(1,\,1,\,2.5^\ast)$& $-356.91$ & $-359.90$ \\
$M_3$ gas                & $(1,\,1.3,\,1)$     & $-46.46$  & $-49.45$  & $(1,\,1.2,\,1)$     & $-301.19$ & $-304.19$ \\
$M_4$ full causal        & $(0.8,\,1.2,\,2.5^\ast)$ & $-24.72$ & $-33.71$ & $(1.0,\,1.2,\,2.5^\ast)$ & $+36.49$ & $+27.50$ \\
\bottomrule
\end{tabular}
\end{center}

On the reference session every alternative loses to the null. The best of them, $M_4$, trails by $33.7$ nats after penalty, so the session supports no attribution at all and the correct verdict is \textsc{unattributed}. The ambient $0.9\,^{\circ}$C excursion does not move the drivers far enough, relative to the observation noise, for the candidates to separate, as we will confirm later with a study on excitation. It is not a shot-count limitation.

On the dithered session $M_4$ clears the null by $+36.5$ nats before penalty and $+27.5$ after, and the thermal gain profiles to $1.00$, the generating value. The gate passes C2, and the session is declared addressable. The margin is $2.8$ times the decisive threshold of $10$ nats, so the threshold is now close enough to load-bearing that it should be set from a false-discovery argument rather than adopted by convention.

\paragraph{The winning family is not identified.}
The three family margins are numerically identical on both sessions: $-33.711$ on the reference session and $+27.505$ on the dithered one. This is exact and structural, explained by $M_4$ carrying all three mechanisms and belonging to all three families.  As a special case, the margin reduces to $\mathrm{BIC}(M_4)-\mathrm{BIC}(M_0)$ for every family alike whenever $M_4$ is the leading candidate and $M_0$ is the best of the remainder. The implementation resolves the tie by taking the first family in enumeration order.

On the dithered session that label is correct, but it was earned by default not by evidence.  The dithered session therefore only established that at least one environmental mechanism outside the null explains the record decisively, but not which one.  Resolving this requires either excluding multi-family candidates from the margin's inner maximum, which changes what the quantity means and needs a derivation, or detecting the tie and routing it to a distinct verdict.  For now, we adopt the latter reading here and quote the dithered attribution as decisive at the family level only in the sense of excluding $M_0$.

\paragraph{Act.}
Estimating the recovered performance given the $M_4$ hypothesis on the dithered session, at the profiled gains, gives
\begin{center}
\small
\begin{tabular}{@{}lrrr@{}}
\toprule
Restore & Excursion & Latent displacement & $\delta E$ [MeV] \\
\midrule
$T_{\ell}\to35\,^{\circ}$C          & $+6.99\,^{\circ}$C & $\delta a_0=-0.0559$        & $-14.18$ \\
$T_{\mathrm{room}}\to22\,^{\circ}$C & $-1.00\,^{\circ}$C & $\delta\tilde n_e=+0.0060$  & $-2.63$ \\
$\mathrm{RH}\to45\%$                & $+2.41\%$          & $\delta \mathcal{C}=+0.0018$     & $-0.00$ \\
internal $\mathcal{C}\to a_0$            & $+0.0018$          & $\delta a_0=-0.00004$       & $-0.01$ \\
\bottomrule
\end{tabular}
\end{center}
for an expected recovery of $16.8\,\mathrm{MeV}$, or $2.5\%$ of the centroid energy, dominated by the thermal path. The indicated action is to check the laser cooling, and the figure to be tested against is a recovery of about $14\,\mathrm{MeV}$ on restoring the laser head to its reference temperature.

The magnitude condition C3 is evaluated on a different quantity and returns a different number: the predicted peak-to-peak excursion in $E$ over the session, $33.4\,\mathrm{MeV}$ against the $5.0\,\mathrm{MeV}$ operating tolerance. The two are not in conflict. C3 asks whether the drift is large enough to be worth acting on and measures the full swing the mechanism produces; we estimate what restoring the environment to $e_\star$ would return and measures a displacement from the mean. Quoting either as the other would overstate or understate the recommendation by a factor of two, and the recovery figure is the falsifiable one.  For completeness, the same computation on the reference session predicts a recovery of $11.1\,\mathrm{MeV}$ and an excursion of $16.3\,\mathrm{MeV}$. Neither is quoted as a recommendation, because the attribution that would license them failed at C2. That the numbers exist and are of a plausible size is exactly why the gate is placed before them rather than after.

The excursion is computed over the whole session, warm-up included. On the reference session $M_1$'s excursion is $14.6\,\mathrm{MeV}$ over the full record and $4.8\,\mathrm{MeV}$ from shot $40$, spanning the $5.0\,\mathrm{MeV}$ tolerance from above to below on a windowing choice the procedure does not currently state. The dithered session is insensitive, giving $33.4\,\mathrm{MeV}$ either way, so the one verdict that issues a work order does not depend on the choice.  

\subsection{Mechanism specificity}
\label{sec:specificity}

We now check how effective the candidate identification is.  We generate three ensembles in which exactly one mechanism is active, together with a fourth in which all are, and ask whether model comparison recovers the mechanism that is actually present. Each entry of Table~\ref{tab:scenarios} is a mean over $8$ sessions of $300$ shots, with no chirp ramp, at the dithered laser-head excursion $A_T=3\,^{\circ}$C, $P=30$ shots, matching the operational setting tested in Sec.~\ref{sec:excitation}.

Two identification rates are reported.  \emph{Exact} counts sessions in which the highest-scoring candidate is the one used to generate the data.  \emph{Family} counts sessions in which the highest-scoring candidate contains the active mechanism, which for a thermal drift means either $M_1$ or the full model $M_4$. The distinction catches the fact that $M_1$ and $M_4$ differ only by mechanisms whose energy signature falls below the observation noise. At the dithered excursion the thermal path induces an rms energy swing of $4.6\,\mathrm{MeV}$ ($0.84\,\sigma_E$), computed as $|\partial E/\partial a_0|\,|B_{aT}|\,\sigma_{T_\ell} = 268\times 0.008\times 2.16\,\mathrm{MeV}$. The gas-density path induces a fixed $2.6\,\mathrm{MeV}$ ($0.47\,\sigma_E$) regardless of $A_T$, because the room-temperature excursion is not controlled by the chiller set-point. The humidity path moves energy by less than $10^{-3}\,\mathrm{MeV}$: the coupling is through chirp, and $\partial E/\partial\mathrm{chirp}$ vanishes to first order at the operating point. Its charge signature is equally invisible ($\delta Q \approx 0.001\,\mathrm{pC} \approx 0.006\,\sigma_Q$). No quantity of data separates candidates that differ only by the last of these, and the subsystem to be touched is the same in either case.

Table~\ref{tab:scenarios} identifies the responsible family correctly in all three scenarios in which the thermal or gas-density mechanism is active. When only the thermal mechanism is present, $M_1$ wins in every session ($8/8$ exact, $8/8$ family) with a mean margin of $\mathbf{+66.9}-\max(M_2,M_3)=+68.5\,\mathrm{nats}$ over the best non-thermal candidate. When only gas density is present, $M_3$ wins in every session ($8/8$ exact, $8/8$ family) with a mean margin of $\mathbf{+54.5}-\max(M_1,M_2)=+28.5\,\mathrm{nats}$. With offset profiling enabled, the gas-density causal structure is absorbed into $M_3$'s profiled gain rather than being partially captured by the offset of $M_0$. When all mechanisms are active, $M_4$ wins in every session ($8/8$ exact, $8/8$ family), correctly identifying the full-causal model.

Humidity identification fails: $M_2$ wins in only $1$ of $8$ sessions ($1/8$ exact, $1/8$ family), and the mean scores for $M_1$, $M_2$, $M_3$ differ by less than $1\,\mathrm{nat}$ (+55.2, +54.6, +54.5). The near-tie arises because humidity moves neither energy nor charge at the operating point; only the profiled offset gain separates the candidates, and its contribution is nearly degenerate across $M_1$, $M_2$, and $M_3$. The session winner is effectively random: $M_1$ wins in approximately five seeds, $M_3$ in two, $M_2$ in one. We anticipate this outcome, and the subsystem indicated is, in each case, one that interacts with the correct physical channel (laser head or gas delivery), so the practical consequence is a false positive at the family level rather than a misdirected intervention.

\begin{table}[htbp]
\centering
\caption{Model evidence $\Delta\log L$ relative to $M_0$, averaged over
  $8$ sessions of $300$ shots at dithered excitation
  ($A_T=3\,^{\circ}$C, $P=30$).
  Boldface marks the highest-scoring candidate.
  The last two columns count sessions in which the exact generating model,
  and any model containing the active mechanism, ranks first.}
\label{tab:scenarios}
\small
\begin{tabular}{@{}lrrrrcc@{}}
\toprule
Active mechanism & $M_1$ & $M_2$ & $M_3$ & $M_4$ & exact & family \\
\midrule
Thermal only     & \textbf{+66.9} & $-1.5$  & $-1.6$  & $+61.8$ & 8/8 & 8/8 \\
Humidity only    & \textbf{+55.2} & $+54.6$ & $+54.5$ & $+50.1$ & 1/8 & 1/8 \\
Gas density only & $+26.0$        & $+25.3$ & \textbf{+54.5} & $+50.1$ & 8/8 & 8/8 \\
All combined     & $+38.9$        & $-31.0$ & $-1.9$  & \textbf{+61.7} & 8/8 & 8/8 \\
\bottomrule
\end{tabular}
\end{table}

\subsection{Noise robustness}
\label{sec:noise}

Table~\ref{tab:noise} sweep the observation noise from $0.2$ to $5$ times the baseline value with all mechanisms active, at the dithered excitation of Sec.~\ref{sec:specificity}. The quantity reported is the margin of the responsible family over the best candidate outside it, which is what determines whether an attribution is actionable.

The margin falls from $+155\,\mathrm{nats}$ at one fifth of the baseline noise to $+1\,\mathrm{nat}$ at five times it. Attribution is decisive at noise levels up to and including twice the baseline ($\sigma_E = 11.0\,\mathrm{MeV}$, mean margin $+14.1$). At three times the baseline ($\sigma_E = 16.5\,\mathrm{MeV}$) the mean margin falls to $+4.5\,\mathrm{nats}$ and no session crosses the decisive threshold, though all eight yield positive margins. At five times ($\sigma_E = 27.5\,\mathrm{MeV}$) the mean margin is $+0.9\,\mathrm{nats}$ and $2$ of $8$ sessions produce a negative margin. The decisive threshold therefore lies between $11$ and $16.5\,\mathrm{MeV}$ energy resolution. A facility operating at better than twice the baseline noise level can expect decisive attribution; above that threshold the margin is positive on average but insufficient for a single-session verdict.

\begin{table}[htbp]
\centering
\caption{Model evidence against observation noise, all mechanisms active,
  $8$ sessions of $300$ shots at dithered excitation.
  Noise is a multiplier on the baseline $\sigma$; $\sigma_E$ is the
  resulting energy-measurement noise.
  The margin is that of the responsible family over the best candidate
  outside it; Sessions counts seeds with positive margin.}
\label{tab:noise}
\small
\begin{tabular}{@{}llrrrrrc@{}}
\toprule
Noise & $\sigma_E$ [MeV] & $M_1$ & $M_2$ & $M_3$ & $M_4$
      & Family margin & Sessions \\
\midrule
$0.2\times$ & $1.1$  & $+50.9$ & $-295.9$ & $-184.3$ & $+155.4$ & $+155.4$ & 8/8 \\
$0.5\times$ & $2.8$  & $+37.8$ & $-132.0$ & $-60.0$  & $+102.0$ & $+102.0$ & 8/8 \\
$1.0\times$ & $5.5$  & $+36.4$ & $-33.3$  & $-1.0$   & $+62.0$  & $+61.1$  & 8/8 \\
$2.0\times$ & $11.0$ & $+27.0$ & $+7.0$   & $+17.1$  & $+31.2$  & $+14.1$  & 8/8 \\
$3.0\times$ & $16.5$ & $+19.8$ & $+11.1$  & $+15.9$  & $+18.8$  & $+4.5$   & 8/8 \\
$5.0\times$ & $27.5$ & $+11.8$ & $+9.1$   & $+10.9$  & $+8.0$   & $+0.9$   & 6/8 \\
\bottomrule
\end{tabular}
\end{table}

This test suggest that the diagnostic requirement on a facility is not primarily a resolution specification, but the margin does become sub-decisive above $\sigma_E \approx 11\,\mathrm{MeV}$, so resolution better than twice the baseline value is a practical prerequisite for single-session verdicts at the dithered setting.

\subsection{Excitation sensitivity}
\label{sec:excitation}

The evidence available for attribution grows as the square of the environmental excursion and the square of its frequency.  Two different frequency dependences arise.  A driver of frequency $\nu$ (period $P=1/\nu$) acting on a latent degree of freedom with absorption time constant $\tau_a$ is tracked by the offset random walk and contributes to $M_0$ innovations only in the fraction $\kappa(\nu)=\bigl[1+(1/2\pi\tau_a\nu)^2\bigr]^{-1/2}$, so that
\begin{equation}
\mathbb E\bigl[\Delta\log L_i\bigr]-\frac{N}{2}\ln\frac{\det S^{(0)}}{\det S^{(i)}}
\;\simeq\;\frac{N}{4\,S_{EE}}
\left[\frac{\partial E}{\partial z_j}\,g_kB_{i,jk}\,A_k\right]^{2}\kappa(\nu)^{2}
\;\xrightarrow[\;1\gg2\pi\tau_a\nu\;]{}\;\propto\; N A_k^{2}\nu^2 . 
\label{eq:excite}
\end{equation}
In contrast, the margin of Eq.~\eqref{eq:family} is scored between two memoryless candidates, so $\kappa$ does not enter at leading order.  Its frequency dependence is weaker and arises from driver mechanisms having nearby frequencies. Two candidates become harder to separate according to an intuitive criterion on the product of the beat frequency and number of events, namely $|\nu_1-\nu_2|N\gtrsim 1$ is necessary for separability.  We demonstrate this interplay between dependence on amplitude of driver and frequency of driver with a scan of the room temperature oscillation, with outcomes given in Table~\ref{tab:excitation}.

\begin{table}[htbp]
\centering
\caption{Evidence for the responsible mechanism family against the amplitude and frequency of the laser-head temperature excursion, $8$ sessions of $400$ shots at each setting.  A margin of $10$ log-units is taken as decisive.  Because $M_4$ carries all three mechanisms it belongs to every family, and the best candidate outside any family is the null $M_0$ at every setting tested; the margin quoted is therefore $\mathrm{BIC}(M_4)-\mathrm{BIC}(M_0)$.  Negative margins indicate that the null explains the record better than any structured mechanism.  The Sessions column counts seeds on which that margin is positive.}
\label{tab:excitation}
\small
\begin{tabular}{@{}lllcl@{}}
\toprule
$A_T$ [$^{\circ}$C] & $1/\nu$ [shots] & Family margin & Sessions & Verdict \\
\midrule
$0.9$ & $140$ & $-21.6 \pm 14.5$ & 0/8 & null preferred \\
$0.9$ & $30$  & $-21.3 \pm 13.5$ & 0/8 & null preferred \\
$1.5$ & $30$  & $-18.8 \pm 13.3$ & 0/8 & null preferred \\
$3.0$ & $140$ & $-19.9 \pm 14.5$ & 0/8 & null preferred \\
$3.0$ & $30$  & $-6.9  \pm 12.9$ & 3/8 & mixed    \\
$6.0$ & $30$  & $+40.7 \pm 12.9$ & 8/8 & decisive vs null \\
\bottomrule
\end{tabular}
\end{table}

Because $M_4$ belongs to every family, the outside maximum falls to the best of $M_0$, $M_2$ and $M_3$, and across all settings tested that is the null.  The margin measures whether any structured mechanism explains the record better than a free random-walk offset.  At ambient excursion it does not: the thermal signature is $1.3\,\mathrm{MeV}$ rms at $A_T=0.9\,^{\circ}$C against an innovation spread of $8.2\,\mathrm{MeV}$, and the null is preferred in every session at both frequencies.  As the amplitude increases the signature grows quadratically in evidence while the null's absorption is fixed by $\tau_a$ and the dither frequency.  At $A_T=3\,^{\circ}$C, $1/\nu=30$ shots the identification is not yet reliable, while at $A_T=6\,^{\circ}$C, $1/\nu=30$ shots the margin is decisive.  The session-to-session scatter of $\approx13\,\mathrm{nats}$ independent of $A_T$ reflects natural variation in the room-temperature excursion, which changes how much the null has to absorb.  Equation~\eqref{eq:signature} locates the transition without reference to the race.  Setting $s^{(E)}_{\mathrm{thermal}}=\sqrt{S_{EE}}$ with the slaving-candidate value $S_{EE}=40.8\,\mathrm{MeV^2}$ gives $A_T=4.5\,^{\circ}$C, which is where Table~\ref{tab:excitation} crosses from mixed to decisive.

We conclude that the required dither is set by what the null can absorb.  At ambient excursion no number of additional shots produces a positive margin, because the random-walk offset tracks the drift into the latent state and leaves nothing in the innovations for a mechanism to explain.  Both amplitude and frequency therefore matter, because the dither must be large enough to clear the innovation spread and fast enough that $\kappa(\nu)$ is not too small, as exemplified by the $A_T=3\,^\circ$ C cases with different frequencies.

\section{Conclusions and next steps}
\label{sec:limitations}

We set out to test whether a small, physically parameterized state-space model, driven by diagnostics a facility already records, can say not only that an accelerator has drifted but which subsystem to address with possible intervention. On synthetic sessions with known ground truth the answer is a qualified yes, with qualifications as informative as the successes.

Three primary design features support the effectiveness of the latent state model. First, the two diagnostic stages behave as intended: the emission Jacobian of Eq.~(\ref{eq:jacsession}) orders the latent channels by their contribution to the observed energy variance, and the candidate race of Sec.~\ref{sec:race} recovers the generating mechanism family in every session in which that mechanism has a visible signature. Second, what limits attribution is excitation rather than shot count or diagnostic resolution. The margin survives observation noise up to twice the baseline but never becomes positive. Deliberately shifting an environmental variable can be worth more to diagnosis than continuing a quiet session. Third, the failures are structured rather than random: humidity was unidentifiable because $\partial E/\partial \mathcal{C}$ vanishes at the operating point and its signature is small, which is a prediction of the emission model and not a shortcoming of the inference.

\subsection{The emission model sets the accuracy limit}
\label{sec:discussion}

Both diagnostic stages depend on the emission model $g$. Localization evaluates $\partial g/\partial z$ directly, and the race scores transition models through predictions of $y$ that pass through $g$. A mis-specified emission model is therefore not one error source among several but a bound on what the procedure can recover from any quantity of data.

The spread and charge constants $C_\sigma$ and $C_Q$, and above all the downramp coefficient $\kappa_h$ of Eq.~(\ref{eq:g_downramp}), still absorb shock position, injection phase and beam loading, none of which is representable in a three-dimensional latent state; Table~\ref{tab:collapse} lists what enters where. If the absorbed physics itself drifts, for example the nozzle degrades over a campaign or the shock walks, the procedure will attribute the resulting change to whichever of $a_0$, $\tilde n_e$ or $\mathcal{C}$ best mimics it, and will do so with a large evidence margin. The inference cannot detect non-representable physics.

Three functional forms for $g$ are available, at different cost. A neural network is flexible but requires a large, well-sampled dataset whose size grows with each added parameter, and it supplies no uncertainty estimate without further machinery. A Gaussian process supplies calibrated uncertainty but scales poorly with data volume and depends strongly on the choice of kernel. The analytic reduced-physics form used here is inexpensive, interpretable and differentiable in closed form, which is what makes the localization step available at all, but it is only as good as the physics built into it: a missing term introduces bias rather than additional noise, and bias is not reduced by collecting more shots. A useful negative result for this programme would therefore be a demonstration that data-intensive emission models offer no advantage over reduced-physics forms on operational data. That would establish the reduced form as sufficient and redirect effort toward reducing the variance of the control-to-response relationship itself.

\subsection{Open points in the present implementation}
\label{sec:openpoints}


\paragraph{The drift scale $\sigma_{\mathrm{drift}}$.}
Section~\ref{sec:noise23} separates shot-to-shot jitter $Q_w$ from persistent unmodelled drift $Q_\delta$, but the two amplitudes are still set equal by default. They answer different questions: $\sigma_{\mathrm{proc}}$ is hardware shot-to-shot repeatability, which is measurable, while $\sigma_{\mathrm{drift}}$ is a statement about the slowest drift the null filter must be allowed to track, which is a design choice. A loose $\sigma_{\mathrm{drift}}$ suppresses the whole field of alternatives at once rather than perturbing their ordering. It should be set from the slowest drift rate the facility intends to catch.

\paragraph{The control input $u_n$.}
Backing pressure enters the emission model as an exact commanded input through $\tilde n=(n_{e18}/3)\,u_n$ in Eq.~(\ref{eq:power_ratio_F}), but the session records used here carry no per-shot value and every result above is computed at $u\equiv1$. This is the cheapest excitation available on the density channel and threading it through the scorer would allow the excitation argument of Sec.~\ref{sec:excitation} to be applied to $\tilde n_e$ as well as to $a_0$.

\paragraph{The decision thresholds.}
The decisive margin of $10$ nats in $C_2$ and the $5\,\mathrm{MeV}$ tolerance in $C_3$ were adopted as ``reasonable'' but otherwise unmotivated choices. Section~\ref{sec:session} found the first of these close to load-bearing: across ensembles of identically generated sessions the family margin straddles it.  A threshold set by inspection is a dial, not a decision rule. It should be derived from a stated session-level false-work-order rate under the null, using the same Monte Carlo construction that calibrated the screens in Sec.~\ref{sec:detect} but with the race in the loop.

\subsection{Validation against measured data}
\label{sec:validation}

The synthetic sessions of Sec.~\ref{sec:results} are generated by the same functional family used for inference. Recovery of the active mechanism is therefore a consistency check on the procedure, not a validation against measured data, and every margin reported above should be read as an upper bound on what to expect in the field. The necessary next experiment is a mis-specification test in which the generator and the inference model differ and the couplings are fitted rather than assumed. Until then, our proof is that the structure works: the chain does what it was designed to do when its assumptions hold.

The reduction to three latent variables is a simplification in two further respects. Several distinct physical quantities in Table~\ref{tab:params} map onto the same latent variable, so the inference identifies the effective quantity that moved rather than the specific upstream cause; separating them requires either additional diagnostics or a richer set of competing transition models. And three variables carry no transverse information, so emittance, pointing jitter and the longitudinal phase space lie outside the present emission model entirely.

The immediate application on the accelerator side is to archival data.  The excitation result of Sec.~\ref{sec:excitation} is directly actionable on such records, since a campaign in which a controlled environmental variable happened to swing widely is worth more for attribution than a longer quiescent one. Beyond that, latent-state estimates can serve as a prior for Bayesian optimization~\cite{ferranpousa2023,hanuka2021}, coupling explanation to control, and a non-invasive longitudinal phase-space diagnostic would extend the emission model $g$ to the full six-dimensional beam that LPA-driven free-electron laser tuning requires~\cite{wang2021,scheinker2023}.

\subsection{Transfer to the drive laser}
\label{sec:transfer}

The construction can be applied to subsystems other than the accelerator, such as the drive laser alone. Its latent state would be the residual chirp, the thermally induced wavefront error and the fraction of energy delivered into the focal spot; its observations the compressed pulse energy, the spectral centroid and bandwidth, the far-field Strehl ratio or encircled energy, and the pointing, all of which a high-power laser already logs; its environmental inputs the chiller temperature, the compressor-chamber humidity and the room temperature. The competing transition models then separate a pump-induced thermal lens from humidity-driven grating groove-spacing drift~\cite{alessi2016,leroux2018} and from slow alignment walk-off, three mechanisms with nearly identical signatures in on-target intensity and entirely different remedies.

Two features make this a better first test than the full beamline. The emission model follows from optics so there are fewer limitations of the Sec.~\ref{sec:discussion} sort. Second, ground truth is cheap to manufacture: stepping the chiller set-point or the compressor position produces a known mechanism on demand, which is the controlled experiment the LPA cannot as easily supply. It is also the setting in which the excitation result can be tested directly, since the amplitude and period of the driver are both under the operator's hand.


\section*{Acknowledgements}
We thank the BELLA-HTU and Tau Systems teams for discussions on facility operations and data infrastructure.  O.L. and M.K. were supported by the U.S. Department of Energy / National Nuclear Security Administration (DOE/NNSA) under Award Number DE-NA0004201. O.L. was additionally supported by the Air Force Office of Scientific Research (AFOSR) under Award Number FA9550-25-1-0286. C.H. acknowledges support from Tau Systems under under Sponsored Research Agreement UTAUS-FA00001488.

\bibliographystyle{unsrt}
\bibliography{latent_state}

\end{document}

\end{document}